\documentclass[11pt]{article}
\pdfoutput=1

\usepackage{jheppub}
\bibstyle{JHEP}

\usepackage[english]{babel}
\usepackage{physics}
\usepackage{tikz}

\usepackage[colorlinks=true,urlcolor=blue,linkcolor=blue,citecolor=blue,linktocpage=true]{hyperref}

\newcommand{\be}{\begin{equation}}
\newcommand{\ee}{\end{equation}}
\newcommand{\ba}{\begin{array}}
\newcommand{\ea}{\end{array}}
\newcommand{\bea}{\begin{equation}\begin{aligned}}
\newcommand{\eea}{\end{aligned}\end{equation}}

\newcommand{\cO}{\mathcal{O}}
\newcommand{\cH}{\mathcal{H}}
\newcommand{\cM}{\mathcal{M}}
\newcommand{\cS}{\mathcal{S}}
\newcommand{\NN}{\mathbb{N}}
\newcommand{\ZZ}{\mathbb{Z}}
\newcommand{\CC}{\mathbb{C}}
\newcommand{\PSLtwoC}{{\mathrm{PSL}(2,\mathbb{C})}}
\newcommand{\secref}[1]{{sec.~\ref{#1}}}
\newcommand{\figref}[1]{{fig.~\ref{#1}}}

\title{Bulk quantum corrections to entwinement}

\author[a]{Marius Gerbershagen,}
\author[a]{Dongming He}
\affiliation[a]{ Theoretische Natuurkunde, Vrije Universiteit Brussel (VUB) and The International Solvay Institutes, Pleinlaan 2, B-1050 Brussels, Belgium}
\emailAdd{marius.gerbershagen@vub.be}
\emailAdd{dongming.he@vub.be}
\abstract{We determine $1/N$ corrections to a notion of generalized entanglement entropy known as entwinement dual to the length of a winding geodesic in asymptotically AdS$_3$ geometries. We explain how $1/N$ corrections can be computed formally via the FLM formula by relating entwinement to an ordinary entanglement entropy in a fictitious covering space. Moreover, we explicitly compute $1/N$ corrections to entwinement for thermal states and small winding numbers using a monodromy method to determine the corrections to the dominant conformal block for the replica partition function. We also determine a set of universal corrections at finite temperature for large winding numbers. Finally, we discuss the implications of our results for the ``entanglement builds geometry'' proposal.}

\begin{document}

\maketitle

\section{Introduction}
The nature of the holographic mapping underlying the AdS/CFT correspondence encodes many local features of physics in the AdS space in a non-local way in the boundary CFT.
In particular, even the geometry of the AdS space itself seems to be encoded in terms of entanglement in the boundary CFT \cite{Swingle:2009bg,VanRaamsdonk:2010pw}.
The main indication for this comes from the fact that the entanglement entropy of a spatial subregion $A$ in a holographic CFT aquires a simple geometric dual in the AdS space.
Namely, to leading order in an expansion in $G_N$ the entanglement entropy is dual to the area of the minimal codimension-two surface $\gamma_A$ homologous to the boundary subregion \cite{Ryu:2006bv}.
The first subleading order is given by the entanglement entropy of the bulk fields in the subregion of the AdS space lying between the RT surface $\gamma_A$ and the asymptotic boundary (the so-called entanglement wedge) \cite{Faulkner:2013ana},
\begin{equation}
  S_A = \frac{\gamma_A}{4G_N} + S_\text{bulk} + O(G_N).
  \label{eq:FLM}
\end{equation}
Eq.~\eqref{eq:FLM} is known as the Faulkner-Lewkowycz-Maldacena (FLM) formula.
Moreover, a so-called quantum extremal surface (QES) prescription has been introduced which captures all orders in the $G_N$ expansion \cite{Engelhardt:2014gca}.

A feature that all of the aformentioned formulas have in common is that they consider entanglement only between spatial degrees of freedom, that is between all of the degrees of freedom in the subregion $A$ and all of the degrees of freedom in the complement of $A$.
But holographic CFTs generically have a large number of fields in the semiclassical limit and thus a large number of internal degrees of freedom that are not spatially organized.
Therefore, in light of the known results for the ordinary spatial entanglement entropy, it is interesting to study entanglement between different fields and ask if the corresponding entanglement entropy has a geometric dual in the AdS space.
For AdS$_3$/CFT$_2$, this question can be answered affirmatively.
There exist particular bipartitions that split both the internal as well as the spatial degrees of freedom into two parts such that the corresponding entanglement entropy is dual to the area of a locally but not globally minimal codimension two surface $\gamma_{A,w}$ \cite{Balasubramanian:2014sra,Balasubramanian:2016xho,Balasubramanian:2018ajb,Erdmenger:2019lzr,Gerbershagen:2021gvc,Craps:2022pke}.
This surface $\gamma_{A,w}$ has non-zero winding number $w>0$ around a conical defect or black hole horizon in the AdS space.
This kind of entanglement is also called \emph{entwinement}.
From the perspective of the ``entanglement builds geometry'' idea, the geodesics with non-zero winding numbers are quite interesting since they probe deeper into the bulk than non-winding geodesics.
In particular, there are finite size regions called \emph{entanglement shadows} around naked singularities and horizons (for one-sided black holes) or singularities (for two-sided black holes) of BTZ black holes which no RT surface enters \cite{Czech:2012bh,Hubeny:2013gta,Nogueira:2013if,Balasubramanian:2014sra,Freivogel:2014lja}.
The entanglement shadows are probed by geodesics with non-zero winding number.
Interestingly, it is precisely these regions close to spacetime singularities where one might expect the largest quantum gravity effects to occur.
Thus, it is particular interesting to study higher order terms in the large $N$ expansion, i.e.~large central charge expansion, for the entanglement between non-spatial degrees of freedom.
From the bulk perspective, these higher order terms represent quantum corrections, i.e.~higher order terms in $G_N$.

With this motivation in mind, in this paper we proceed to study the leading $1/N$ corrections to entwinement for states dual to BTZ black holes and conical defects.
In \secref{sec:review}, we review some aspects of entwinement relevant for our work.
Sec.~\ref{sec:covering-space} proceeds to explain how to relate entwinement to an ordinary entanglement entropy in a fictitious covering theory as well as how to apply the FLM formula \cite{Faulkner:2013ana} to this system.
In \secref{sec:monodromy-method}, we explicitly compute $1/N$ corrections to entwinement by a replica trick, using a monodromy method to compute the $1/N$ corrections to the dominant conformal block for the replica partition function.
These results apply for small winding numbers.
For large winding numbers, universal finite size and finite temperature corrections to entwinement at finite temperature are computed in \secref{sec:large-winding-numbers}.
Finally, \secref{sec:discussion} contains a discussion of our results, in particular their implications for the ``entanglement builds geometry'' idea.

\section{Review of aspects of entwinement}
\label{sec:review}
Let us first review the most important aspects of the notion of non-spatial entanglement we are going to use in the rest of this publication.
The study of ordinary spatial entanglement in quantum field theories is based on a bipartition of the degrees of freedom into those associated to a spatial subregion $A$ and those associated to the complement $A^c$.
In that case, we are considering a bipartition where we look at all fields of the theory in the same subregion $A$.
The basic idea of non-spatial entanglement is to consider different fields $X_i$ in different subregions $A_i$ in order to define a bipartition of the degrees of freedom.
The standard definition of the entanglement entropy as the von Neumann entropy of the reduced density matrix given by tracing over a Hilbert space factor applies also to this setup,\footnote{Note that strictly speaking this definition does not apply to gauge theories, even for the ordinary entanglement entropy, as there is no factorization of the Hilbert space into tensor factors \cite{Buividovich:2008gq,Donnelly:2011hn,Casini:2013rba,Radicevic:2014kqa,Aoki:2015bsa,Ghosh:2015iwa,Soni:2015yga}. Since entwinement is defined in gauge theories, this issue applies there as well and formally a more general procedure has to used in order to define entwinement in a gauge-invariant way \cite{Erdmenger:2019lzr}.}
\begin{equation}
    \cH = \cH_{\{A_i\}}\otimes\cH_{\{A_i^c\}}, \quad \rho_{\{A_i\}} = \Tr_{\cH_{\{A_i^c\}}}\rho, \quad S_{\{A_i\}} = \Tr_{\cH_{\{A_i\}}}[\rho_{\{A_i\}}\log\rho_{\{A_i\}}].
\end{equation}
Entwinement as considered below is a special case of this setup where we consider a class of two-dimensional holographic CFTs and choose the subregions $A_i$ in such a way that the corresponding entanglement entropy is dual to the length of a non-minimal geodesic.
We will now explain the features of this class of holographic CFTs and which choice of subregions $A_i$ is appropriate.
In the following, we will concentrate on two examples of states dual to conical defects and black holes.
For AdS$_3$ gravity, these spacetimes are the most relevant geometries that admit geodesics with non-zero winding numbers.

\subsection{Pure states dual to conical defects}
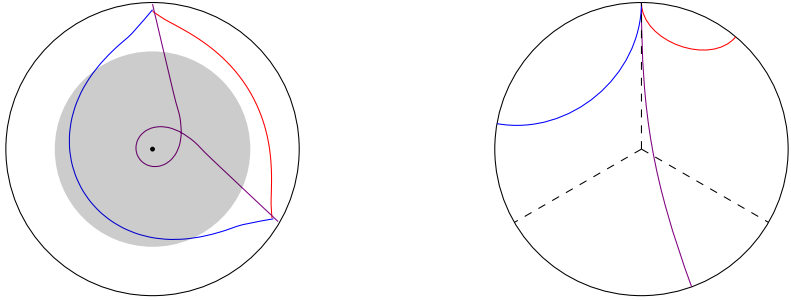
\begin{figure}
  \centering
  \begin{tikzpicture}[scale=0.65]
    \def\i{0}
    \def\radius{3}
    \def\N{3}
    \def\betatilde{60}
    \def\beta{\betatilde/\N}
    \def\xshiftA{-5}
    \begin{scope}[shift={(\xshiftA,0)}]
      \draw (0,0) circle (\radius);

      \def\alpha{min(\delta,180-\delta)}
      \def\signflip{sign(90-\delta)}
      \def\r{atan(sqrt(abs((1+tan(\phi)*tan(\phi))/(tan(\alpha)*tan(\alpha)+1e-2-tan(\phi)*tan(\phi)))))}
      \foreach \j [evaluate=\j as \color using {int((\j-1)*50)}] in {1,...,\N}
      {
        \def\delta{(\beta+180/\N*(\j-1))}
        \draw[color=blue!\color!red] plot[domain=-\alpha:\alpha,smooth,variable=\phi] ({\r*\radius/90*sin((\phi+\signflip*\alpha)*\N)},{\r*\radius/90*cos((\phi+\signflip*\alpha)*\N)});
      }

      \def\delta{\beta}
      \def\phi{0}

      \fill[black,opacity=0.2] (0,0) circle ({{atan(1/tan(90/\N))*\radius/90}});
      \fill[black] (0,0) circle (0.05cm);
    \end{scope}
    
    \def\xshiftB{5}
    \def\scaleB{1}
        
    \begin{scope}[shift={(\xshiftB,0)},scale=\scaleB]
      \draw (0,0) circle (\radius);

      \foreach \j [evaluate=\j as \color using {int((\j-1)*50)}] in {1,...,\N}
      {
        \def\alpha{-2*\beta-2*180/\N*\j+2*180/\N}
        \draw[color=blue!\color!red] ({{\radius*sin(\i*360/\N)}},{{\radius*cos(\i*360/\N)}}) to[out={{-90-\i*360/\N}},in={{-90-\i*360/\N+\alpha}}] ({{-\radius*sin(\i*360/\N+\alpha)}},{{\radius*cos(\i*360/\N+\alpha)}});
        \draw[dashed] (0,0) -- ({{\radius*sin(\j*360/\N)}},{{\radius*cos(\j*360/\N)}});
      }

      \def\eradiusx{\radius/2}
      \def\eradiusy{\radius/4}
      \def\abstand{\radius+3+\eradiusx}
      \def\A{0.5}
    \end{scope}
  \end{tikzpicture}
  \caption{The conical defect geometry (left) is obtained by partitioning a pure AdS$_3$ covering space (right) into $m$ wedges (shown here is the case $m=3$) which are identified with each other.
    Ryu-Takayanagi surfaces, that is geodesics with winding number zero, do not penetrate the entanglement shadow indicated in gray.
    However, the remaining geodesics with winding number $w>0$ shown in blue and purple probe the entire constant time slice of the conical defect.}
  \label{fig:conical defect}
\end{figure}
A conical defect with asymptotically AdS$_3$ boundary is a spacetime with constant negative curvature and a naked conical singularity in the center.
It can be constructed by partitioning a pure AdS$_3$ spacetime (the so-called \emph{covering space}) into $m \in \NN$ wedges that are then identified with each other (see \figref{fig:conical defect}) \cite{Balasubramanian:2014sra}.
We will call the integer $m$ \emph{covering parameter}.
In a bottom-up picture, the dual CFT state is given by the ground state of the twisted sector of a $\ZZ_m$ orbifold \cite{Balasubramanian:2014sra}.
This orbifold theory is obtained by taking $m$ identical copies of a so-called ``seed CFT'', which can be any large central charge holographic CFT.
For each field $X$ in the seed theory there are $m$ identical copies $X_i$ in the orbifold theory which are identified by the $\ZZ_m$ permutations.
Thus, the entire field content of the theory splits into groups of $m$ fields.
In the twisted sector, the fields obey boundary conditions that are periodic up to $\ZZ_m$ permutations, $X_i(\phi+2\pi) = X_{i+1}(\phi)$.
That means that a field and its $m-1$ copies can be put together into one single field with periodic boundary conditions, $\tilde X(\tilde \phi+2\pi) = \tilde X(\tilde\phi)$ where $\tilde\phi = \phi/m$.
This single field can be though of as living on the boundary of the pure AdS$_3$ covering space.

\begin{figure}
  \centering
  \begin{tikzpicture}[style=thick]
    \begin{scope}
      \draw (0,0.55) node[above] {ordinary entanglement};
      \draw ({{cos(-60)*1.5}},{{sin(-60)*0.375}}) arc(-60:-5:1.5cm and 0.375cm);
      \draw[red,densely dashed] ({{cos(-5)*1.5}},{{sin(-5)*0.375}}) arc(-5:60:1.5cm and 0.375cm);
      \draw ({{cos(60)*1.5}},{{sin(60)*0.375}}) arc(60:110:1.5cm and 0.375cm);
      \draw[red,densely dashed] ({{cos(110)*1.5}},{{sin(110)*0.375}}) arc(110:180:1.5cm and 0.375cm);
      \draw ({{cos(180)*1.5}},{{sin(180)*0.375}}) arc(180:250:1.5cm and 0.375cm);
      \draw[red,densely dashed] ({{cos(250)*1.5}},{{sin(250)*0.375}}) arc(250:300:1.5cm and 0.375cm);
      \draw ({{cos(60)*1.5-0.12}},{{-sin(60)*0.375+0.07}}) -- ({{cos(60)*1.5+0.12}},{{-sin(60)*0.375-0.07}});
      \draw (-1.65,0) -- (-1.35,0);
      \draw ({{cos(60)*1.5-0.12}},{{sin(60)*0.375-0.07}}) -- ({{cos(60)*1.5+0.12}},{{sin(60)*0.375+0.07}});
      \fill[color=red] ({{cos(-5)*1.5}},{{sin(-5)*0.375}}) circle(0.05);
      \fill[color=red] ({{cos(60)*1.5}},{{sin(60)*0.375}}) circle(0.05);
      \fill[color=red] ({{cos(110)*1.5}},{{sin(110)*0.375}}) circle(0.05);
      \fill[color=red] ({{cos(180)*1.5}},{{sin(180)*0.375}}) circle(0.05);
      \fill[color=red] ({{cos(250)*1.5}},{{sin(250)*0.375}}) circle(0.05);
      \fill[color=red] ({{cos(300)*1.5}},{{sin(300)*0.375}}) circle(0.05);
    \end{scope}
    \draw (2.3,-0.9) -- (2.3,1);
    \draw (-2,0.5) -- (13,0.5);
    \begin{scope}[shift={(8,0)}]
      \draw (0,0.55) node[above] {entwinement};
      \begin{scope}[shift={(-3.5,0)}]
        \draw[red,densely dashed] ({{cos(110)*1.5}},{{sin(110)*0.375}}) arc(110:180:1.5cm and 0.375cm);
        \draw ({{cos(180)*1.5}},{{sin(180)*0.375}}) arc(180:470:1.5cm and 0.375cm);
        \draw ({{cos(60)*1.5-0.12}},{{-sin(60)*0.375+0.07}}) -- ({{cos(60)*1.5+0.12}},{{-sin(60)*0.375-0.06}});
        \draw (-1.65,0) -- (-1.35,0);
        \draw ({{cos(60)*1.5-0.12}},{{sin(60)*0.375-0.07}}) -- ({{cos(60)*1.5+0.12}},{{sin(60)*0.375+0.06}});
        \fill[color=red] ({{cos(110)*1.5}},{{sin(110)*0.375}}) circle(0.05);
        \fill[color=red] ({{cos(180)*1.5}},{{sin(180)*0.375}}) circle(0.05);
        \draw (0,-0.4) node[below] {$w=0$};
      \end{scope}
      \begin{scope}
        \draw[blue!50!red,densely dashed] ({{cos(10)*1.5}},{{-sin(10)*0.375}}) arc(-10:180:1.5cm and 0.375cm);
        \draw ({{cos(180)*1.5}},{{sin(180)*0.375}}) arc(180:355:1.5cm and 0.375cm);
        \draw ({{cos(60)*1.5-0.12}},{{-sin(60)*0.375+0.07}}) -- ({{cos(60)*1.5+0.12}},{{-sin(60)*0.375-0.07}});
        \draw (-1.65,0) -- (-1.35,0);
        \draw ({{cos(60)*1.5-0.12}},{{sin(60)*0.375-0.07}}) -- ({{cos(60)*1.5+0.12}},{{sin(60)*0.375+0.07}});
        \fill[color=blue!50!red] ({{cos(10)*1.5}},{{-sin(10)*0.375}}) circle(0.05);
        \fill[color=blue!50!red] ({{cos(180)*1.5}},{{sin(180)*0.375}}) circle(0.05);
        \draw (0,-0.4) node[below] {$w=1$};
      \end{scope}
      \begin{scope}[shift={(3.5,0)}]
        \draw[blue,densely dashed] ({{cos(110)*1.5}},{{-sin(110)*0.375}}) arc(-110:180:1.5cm and 0.375cm);
        \draw ({{cos(180)*1.5}},{{sin(180)*0.375}}) arc(180:250:1.5cm and 0.375cm);
        \draw ({{cos(60)*1.5-0.12}},{{-sin(60)*0.375+0.07}}) -- ({{cos(60)*1.5+0.12}},{{-sin(60)*0.375-0.07}});
        \draw (-1.65,0) -- (-1.35,0);
        \draw ({{cos(60)*1.5-0.12}},{{sin(60)*0.375-0.07}}) -- ({{cos(60)*1.5+0.12}},{{sin(60)*0.375+0.07}});
        \fill[color=blue] ({{cos(110)*1.5}},{{-sin(110)*0.375}}) circle(0.05);
        \fill[color=blue] ({{cos(180)*1.5}},{{sin(180)*0.375}}) circle(0.05);
        \draw (0,-0.4) node[below] {$w=2$};
      \end{scope}
    \end{scope}
  \end{tikzpicture}
  \caption{Degrees of freedom on the covering space of the conical defect for ordinary entanglement and for entwinement (shown here is the case $m=3$).
    The dashed lines correspond to the subsystem whose entanglement is computed.}
  \label{fig:DoF entwinement conical defect}
\end{figure}
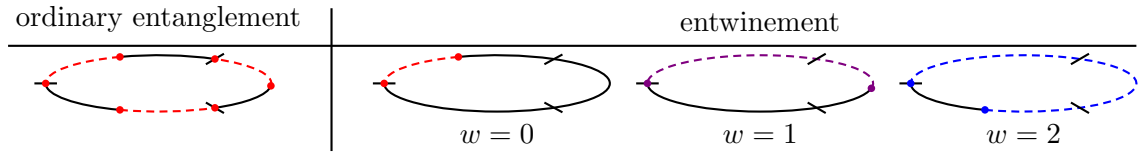
As mentioned above, for entwinement we will consider a subset of the degrees of freedom where different fields are localized in different subregions \cite{Balasubramanian:2014sra}.
More precisely, we will choose a bipartition where we consider $w < m$ fields in the entire space together with one field in an interval $[0,L]$ covering only a part of the spatial direction (see \figref{fig:DoF entwinement conical defect}).
The integer $w$ is called \emph{winding number}.
The parameter $L$ is restricted to $0<L<1$.
This bipartition corresponds to considering a single interval $\tilde\phi \in [0,(w+L)/m]$ on the boundary of the covering space (note that in our conventions, the periodic spatial direction has length one).
The entanglement entropy for this bipartition is dual to the length of a geodesic with winding number $w$ and opening angle $2\pi L$ in the conical defect \cite{Balasubramanian:2014sra},
\begin{equation}
  S_{w,m}(L) = \frac{\gamma_{L,w}}{4mG_N} + O(G_N^0).
\end{equation}
The winding geodesic in the conical defect descends from an ordinary geodesic with opening angle $2\pi(w+L)/m$ in the covering space after the identification procedure that leads to the conical defect has been performed.

\subsection{Mixed states at finite temperature}
In AdS/CFT, black holes are dual to thermal states in the boundary CFT.
We will assume that the boundary CFT has the Hilbert space structure of an $S_N$ orbifold theory.\footnote{For string theory constructions in AdS$_3 \times X$ with pure NS-NS flux, this seems to be true generically, see e.g.~\cite{Eberhardt:2021vsx,Knighton:2024qxd,Sriprachyakul:2024gyl}.}
This means that we now consider $N$ copies of a seed CFT identified under an $S_N$ action.
Unlike before, the seed CFT does not need to be holographic.\footnote{As long as the covering parameter $m$ for the conical defect does not scale with the central charge, the seed CFT of the conical defect has to be a large central charge holographic CFT in order for classical gravity in the conical defect background to be a reliable description of the bulk. For the $S_N$ orbifold, the central charge scales with $N$ and the seed CFT does not have to have any holographic features.}
Large $N$ corresponds to large central charge and a small $G_N$ semiclassical limit.
In later sections we will compute the first subleading term in a large $N$ expansion.

As before, we choose a bipartition that localizes different fields in different subregions \cite{Gerbershagen:2021gvc}.
In contrast to the conical defect, however, there is now also a new ingredient in the definition of entwinement: we also restrict to particular boundary conditions for the fields, i.e.~particular twisted sectors of the $S_N$ orbifold.
Unlike for the conical defect dual to a pure state in a single twisted sector, the thermal state is a mixture of states from all possible twisted sectors.
Since these twisted sectors are independent (the total Hilbert space is a direct sum of twisted sector Hilbert spaces, $\cH = \bigoplus_s \cH_s$ and the thermal state is block diagonal), we can choose the subregions for each twisted sector separately.
For entwinement, we only consider non-empty subregions for some subset $\cS_m \subset \cH$ of twisted sectors or in other words, we restrict to a particular set of boundary conditions.

\begin{figure}
  \centering
  \begin{tikzpicture}[scale=1.1,style=thick]
    \begin{scope}
      \begin{scope}
        \draw ({{cos(35)}},{{sin(35)*0.25}}) arc(35:180:1cm and 0.25cm);
        \draw[red,densely dashed] ({{cos(180)}},{{sin(180)*0.25}}) arc(180:395:1cm and 0.25cm);
        \draw (-0.85,0) -- (-1.15,0);
        \fill[color=red] ({{cos(180)}},{{sin(180)*0.25}}) circle(0.05);
        \fill[color=red] ({{cos(35)}},{{sin(35)*0.25}}) circle(0.05);
      \end{scope}
      \begin{scope}[shift={(3.5,0)}]
        \draw ({{cos(295)*1.33}},{{sin(295)*0.33}}) arc(295:360:1.33cm and 0.33cm);
        \draw[red,densely dashed] ({{cos(0)*1.33}},{{sin(0)*0.33}}) arc(0:105:1.33cm and 0.33cm);
        \draw ({{cos(105)*1.33}},{{sin(105)*0.33}}) arc(105:180:1.33cm and 0.33cm);
        \draw[red,densely dashed] ({{cos(180)*1.33}},{{sin(180)*0.33}}) arc(180:295:1.33cm and 0.33cm);
        \draw (-1.48,0) -- (-1.18,0);
        \draw (1.48,0) -- (1.18,0);
        \fill[color=red] (-1.33,0) circle(0.05);
        \fill[color=red] (1.33,0) circle(0.05);
        \fill[color=red] ({{cos(105)*1.33}},{{sin(105)*0.33}}) circle(0.05);
        \fill[color=red] ({{cos(295)*1.33}},{{sin(295)*0.33}}) circle(0.05);
      \end{scope}
      \begin{scope}[shift={(7.5,0)}]
        \draw ({{cos(-60)*1.5}},{{sin(-60)*0.375}}) arc(-60:-5:1.5cm and 0.375cm);
        \draw[red,densely dashed] ({{cos(-5)*1.5}},{{sin(-5)*0.375}}) arc(-5:60:1.5cm and 0.375cm);
        \draw ({{cos(60)*1.5}},{{sin(60)*0.375}}) arc(60:110:1.5cm and 0.375cm);
        \draw[red,densely dashed] ({{cos(110)*1.5}},{{sin(110)*0.375}}) arc(110:180:1.5cm and 0.375cm);
        \draw ({{cos(180)*1.5}},{{sin(180)*0.375}}) arc(180:250:1.5cm and 0.375cm);
        \draw[red,densely dashed] ({{cos(250)*1.5}},{{sin(250)*0.375}}) arc(250:300:1.5cm and 0.375cm);
        \draw ({{cos(60)*1.5-0.12}},{{-sin(60)*0.375+0.07}}) -- ({{cos(60)*1.5+0.12}},{{-sin(60)*0.375-0.07}});
        \draw (-1.65,0) -- (-1.35,0);
        \draw ({{cos(60)*1.5-0.12}},{{sin(60)*0.375-0.07}}) -- ({{cos(60)*1.5+0.12}},{{sin(60)*0.375+0.07}});
        \fill[color=red] ({{cos(-5)*1.5}},{{sin(-5)*0.375}}) circle(0.05);
        \fill[color=red] ({{cos(60)*1.5}},{{sin(60)*0.375}}) circle(0.05);
        \fill[color=red] ({{cos(110)*1.5}},{{sin(110)*0.375}}) circle(0.05);
        \fill[color=red] ({{cos(180)*1.5}},{{sin(180)*0.375}}) circle(0.05);
        \fill[color=red] ({{cos(250)*1.5}},{{sin(250)*0.375}}) circle(0.05);
        \fill[color=red] ({{cos(300)*1.5}},{{sin(300)*0.375}}) circle(0.05);
      \end{scope}
      \draw (10,0) node {\textbf{...}};
      \draw (4.5,0.6) node[above] {ordinary entanglement:};
    \end{scope}
    \begin{scope}[shift={(0,-2)},scale=0.8]
      \begin{scope}
        \draw[blue,densely dashed] ({{cos(10)*1.5}},{{-sin(10)*0.375}}) arc(-10:180:1.5cm and 0.375cm);
        \draw ({{cos(180)*1.5}},{{sin(180)*0.375}}) arc(180:355:1.5cm and 0.375cm);
        \draw ({{cos(60)*1.5-0.12}},{{-sin(60)*0.375+0.07}}) -- ({{cos(60)*1.5+0.12}},{{-sin(60)*0.375-0.07}});
        \draw (-1.65,0) -- (-1.35,0);
        \draw ({{cos(60)*1.5-0.12}},{{sin(60)*0.375-0.07}}) -- ({{cos(60)*1.5+0.12}},{{sin(60)*0.375+0.07}});
        \fill[color=blue] ({{cos(10)*1.5}},{{-sin(10)*0.375}}) circle(0.06);
        \fill[color=blue] ({{cos(180)*1.5}},{{sin(180)*0.375}}) circle(0.06);
        \draw (0,-0.4) node[below] {$w=1$};
      \end{scope}
      \begin{scope}[shift={(5,0)}]
        \draw[blue,densely dashed] ({{cos(85)*2}},{{sin(85)*0.5}}) arc(85:180:2cm and 0.5cm);
        \draw ({{cos(180)*2}},{{sin(180)*0.5}}) arc(180:275:2cm and 0.5cm);
        \draw[blue,densely dashed] ({{cos(275)*2}},{{sin(275)*0.5}}) arc(275:360:2cm and 0.5cm);
        \draw ({{cos(0)*2}},{{sin(0)*0.5}}) arc(0:85:2cm and 0.5cm);
        \draw (-2.15,0) -- (-1.85,0);
        \draw (2.15,0) -- (1.85,0);
        \draw ({{cos(60)*2-0.12}},{{sin(60)*0.5-0.07}}) -- ({{cos(60)*2+0.12}},{{sin(60)*0.5+0.07}});
        \draw ({{cos(120)*2-0.12}},{{sin(120)*0.5+0.07}}) -- ({{cos(120)*2+0.12}},{{sin(120)*0.5-0.07}});
        \draw ({{cos(240)*2-0.12}},{{sin(240)*0.5-0.07}}) -- ({{cos(240)*2+0.12}},{{sin(240)*0.5+0.07}});
        \draw ({{cos(300)*2-0.12}},{{sin(300)*0.5+0.07}}) -- ({{cos(300)*2+0.12}},{{sin(300)*0.5-0.07}});
        \fill[color=blue] ({{cos(85)*2}},{{sin(85)*0.5}}) circle(0.06);
        \fill[color=blue] ({{cos(180)*2}},{{sin(180)*0.5}}) circle(0.06);
        \fill[color=blue] ({{cos(275)*2}},{{sin(275)*0.5}}) circle(0.06);
        \fill[color=blue] ({{cos(360)*2}},{{sin(360)*0.5}}) circle(0.06);
      \end{scope}
      \draw (9,0) node {\textbf{...}};
      \draw (5,0.6) node[above] {entwinement:};
    \end{scope}
  \end{tikzpicture}
  \caption{Degrees of freedom for entwinement at finite temperature. Top: for ordinary entanglement, long strands of all sizes contribute. We compute the entanglement w.r.t.~a subsystem of $k$ intervals on a strand of length $k$. Bottom: for entwinement, only long strands of length $km$ contribute (here we depict $m=3$). We compute the entanglement w.r.t.~a subsystem of $k$ intervals on a strand of length $km$.}
  \label{fig:DoF entwinement thermal state}
\end{figure}
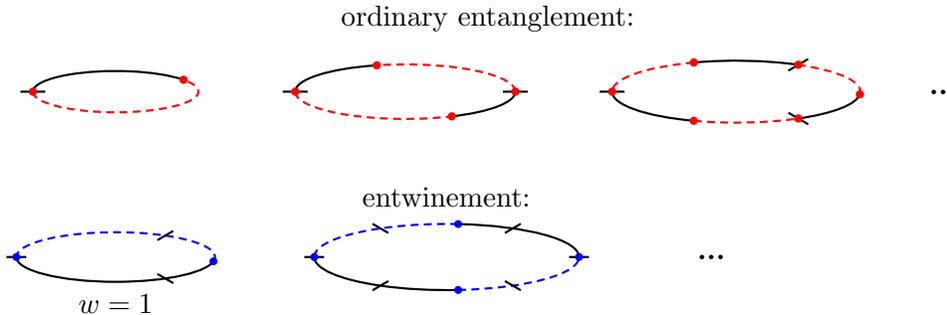

To be precise, we consider twisted sectors where the fields obey boundary conditions that make them periodic up to $S_N$ permutations with cycle lengths that are a multiple of a covering parameter $m \in \NN$.\footnote{Strictly speaking, this is only possible if $N$ is a multiple of $m$. However, for any $N$ one can consider twisted sectors containing as many cycles with length $k m$ ($k \in \NN$) as possible such that the remaining cycles of the twisted sector have total length smaller than $m$. If we choose a vanishing subset of degrees of freedom for the remaining cycles, the result for the entanglement entropy is unaffected by the presence of those cycles. Hence, our results hold for generic values of $N$.}
A cycle that joins $k>1$ fields together through the twisted boundary conditions is also referred to as a \emph{long strand}.
So in other words, for entwinement we are considering only long strands whose length is a multiple of $m$.
Then, for the subset of degrees of freedom, we consider again $w < m$ fields on the entire space along with an interval $[0,L]$ for a further field such that on the long strand, these subregions coalesce into a single interval (see \figref{fig:DoF entwinement thermal state}).
To leading order in the large $N$ expansion, the entanglement entropy for this bipartition is dual to the length of a geodesic in the BTZ spacetime with winding number $w$ and opening angle $2\pi L$ \cite{Gerbershagen:2021gvc} (see \figref{fig:geodesics-BTZ}),
\begin{equation}
  S_{w,m}(L) = \frac{\gamma_{L,w}}{4mG_N} + O(G_N^0).
\end{equation}
As for the ordinary entanglement entropy, there is a phase transition as $w$ increases such that the entanglement entropy for large $w$ is given by the thermal entropy plus the length of a geodesic with winding number of $m-w-1$ and opening angle $2\pi(1-L)$.
Moreover, for small temperatures $\beta > 2\pi m$ the entanglement entropy is dual to the length of a geodesic in thermal AdS$_3$ with opening angle $2\pi(w+L)/m$.
\begin{figure}
  \centering
  \begin{tikzpicture}[scale=0.35]
    \begin{scope}
      \draw (0,0) circle(5);
      \fill[black,opacity=0.3] (0,0) circle(1);
      \fill[black] (0,0) circle(0.5);
      \draw[color=blue] ({{5*cos(60)}},{{5*sin(60)}}) to[out=230,in=0] (0,-0.75) arc(-90:-270:0.75) (0,0.75) to[out=0,in=130] ({{5*cos(60)}},{{-5*sin(60)}});
      \draw[color=red] ({{5*cos(60)}},{{5*sin(60)}}) to[out=230,in=90] (3,0) to[out=-90,in=130] ({{5*cos(60)}},{{-5*sin(60)}});
    \end{scope}
    \begin{scope}[shift={(15,0)}]
      \draw (0,0) circle(5);
      \draw[densely dashed] (0,5) -- (0,0);
      \draw[densely dashed] ({{cos(210)*5}},{{sin(210)*5}}) -- (0,0);
      \draw[densely dashed] ({{cos(330)*5}},{{sin(330)*5}}) -- (0,0);
      \fill[black,opacity=0.3] (0,0) circle(3);
      \fill[black] (0,0) circle(1.5);
      \draw[color=blue] ({{5*cos(90)}},{{5*sin(90)}}) to[out=280,in=135] ({{3.5*cos(45)}},{{3.5*sin(45)}}) arc(45:-45:3.5) ({{3.5*cos(-45)}},{{3.5*sin(-45)}}) to[out=225,in=80] ({{5*cos(-90)}},{{5*sin(-90)}});
      \draw[color=red] ({{5*cos(90)}},{{5*sin(90)}}) to[out=280,in=140] ({{3.75*cos(40)}},{{3.75*sin(40)}}) to[out=-40,in=180] ({{5*cos(10)}},{{5*sin(10)}});
    \end{scope}
  \end{tikzpicture}
  \caption{Winding geodesics in the BTZ black hole background. LHS: geodesics with winding number $w=0$ (in red) and $w=1$ (in blue). RHS: covering space interpretation: the length of geodesics with zero winding number in the covering space BTZ geometry with inverse temperature $\beta/m$ is the same as that of winding geodesics in the original BTZ geometry with inverse temperature $\beta$. Here, we depict $m=3$.}
  \label{fig:geodesics-BTZ}
\end{figure}
At this point the reader may wonder if we are still computing a measure of entanglement if we restrict to particular boundary conditions in the process of doing so.
The answer to this question is yes, although the justification is somewhat technical \cite{Erdmenger:2019lzr}.
The usual entanglement entropy definition assumes a factorizing Hilbert space, $\cH = \cH_A \otimes \cH_{A^c}$.
This factorization property, however, does not hold true in gauge theories such as the $S_N$ gauge theory considered here.
Instead, a gauge invariant way to specify a subsystem is to choose a subset $\cM_A$ of operators that forms the maximal set of possible measurement operators that an observer in the subsystem would have access to.\footnote{For spatial subregions, this subset has an algebra structure. Here we only have a linear subspace which nevertheless suffices to define an entanglement entropy.}
For entwinement, we simply choose a subset of operators $\cM_{m,w}(L)$ that act only within twisted sectors containing solely cycles of length $km$.
That is, when applied to a state $\ket\psi$, an operator $\cO \in \cM_{m,w}(L)$ gives a non-vanishing result only for $\ket\psi \in \cS_m$.
One can associate an entanglement entropy to this subset that qualifies the entropy that an observer with access only to operators from $\cM_{m,w}(L)$ would measure, see \cite{Erdmenger:2019lzr} for details.
Specifying on which twisted sectors the operators under consideration act is necessary in general in order to specify a subset in the $S_N$ orbifold.
Usually, one considers operators acting on all twisted sectors and thus this detail is omitted.
For the conical defect, the dual CFT state is in a single twisted sector and hence the choice of boundary conditions is immaterial as well.

Another important point to make is that in the well-known examples of top-down AdS$_3$/CFT$_2$ constructions such as the D1/D5 system the 2d CFT is typically a deformation of the orbifold theory by an exactly marginal operator (see e.g.~\cite{David:2002wn,Datta:2017ert,Eberhardt:2021vsx,Martinec:2022ofs} for a collection of such constructions).
The undeformed $S_N$ orbifold theory represents a very special point where the CFT is weakly coupled and correspondingly the dual gravity description is strongly coupled.
The deformation breaks the $S_N$ symmetry although the Hilbert space structure as a direct sum of twisted sectors of course continues to be valid.
Since the above definition of entwinement was based on the existence of this Hilbert space structure but not on the fact that the boundary CFT is an exact $S_N$ orbifold theory, the definition of entwinement also applies away from the orbifold point.
Under some reasonable assumptions on the CFT data (sparse spectrum of low dimension operators\footnote{Using the same notion of sparsity as in \cite{Hartman:2014oaa}.} and at most exponentially growing OPE coefficients), the large $N$ result for entwinement is the same at strong and at weak coupling in the CFT \cite{Gerbershagen:2021gvc}.

\section{$1/N$ corrections: covering space picture}
\label{sec:covering-space}
Having reviewed previous work on the leading order result for entwinement in the large $N$ limit, we now turn to the first subleading order. We comment on the interpretation of entwinement as an ordinary entanglement entropy in a covering space and the implications of this interpretation for $1/N$ corrections.
We then discuss the bulk perspective and comment on the applicability of the FLM \cite{Faulkner:2013ana} and QES \cite{Engelhardt:2014gca} formulas to entwinement.

\subsection{Pure states dual to conical defects}
To illustrate the covering space picture, we will start out by explaining the $1/N$ corrections in the simple example of pure states dual to conical defects.
In this case the identification of entwinement with an ordinary entanglement entropy on the covering space is obvious.
The subleading corrections in the large $N$ or large central charge expansion are easily obtained from the known results for $1/N$ corrections to the ordinary entanglement entropy.
The entanglement entropy for a single interval in the vacuum of a 2d CFT is fixed by conformal symmetry and proportional to the central charge.
Thus, there are no $1/N$ corrections for the entwinement dual to the length of a single winding geodesic in the conical defect,
\begin{equation}
  S_{w,m}(L) = \frac{\gamma_{L,w}}{4mG_N} \qquad \forall G_N.
\end{equation}
The $1/m$ prefactor can be understood from the from the covering space picture as follows.
The central charge is a measure for the number of degrees of freedom in the CFT.
By combining $m$ fields into one field in order to form the covering space, the number of degrees of freedom is reduced by a factor of $1/m$, $c \to c/m$.

From the bulk perspective, the $1/N$ corrections to entwinement are given by the FLM formula \cite{Faulkner:2013ana} on the covering space,
\begin{equation}
  S_{w,m}(L) = \frac{\gamma_{L,w}}{4m G_{N,\text{bare}}} + S_\text{bulk} + O(G_{N,\text{bare}}).
\end{equation}
From the bulk side, the leading $1/N$ correction is given by the entanglement wedge entropy $S_\text{bulk}$.
Due to the universality of the single interval entanglement entropy in 2d CFTs, this term must be proportional to $\gamma_{L,w}$ and only renormalizes $G_N$.
Note the factor of $\frac{1}{4m G_{N,\text{bare}}}$ instead of $\frac{1}{4G_{N,\text{bare}}}$ in the leading order.
This change in Newton's constant $G_N \to m G_N$ on the covering space follows from the aforementioned reduction in the number of degrees of freedom, $c \to c/m$, and the well-known relation $c=\frac{3l}{2G_N}$ between central charge and Newton's constant.

\subsection{Mixed states at finite temperature}
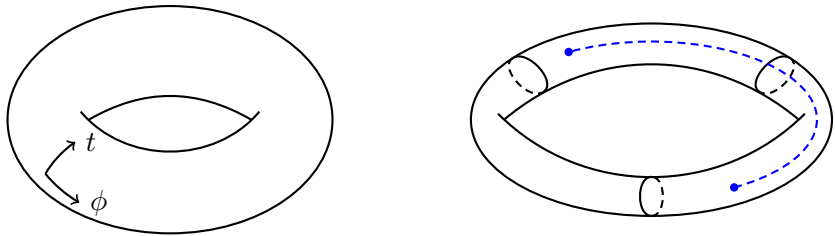
\begin{figure}
  \centering
  \begin{tikzpicture}[scale=0.8,style=thick]
    \begin{scope}[scale=0.9]
      \draw (0,0) ellipse (3cm and 2.1cm);
      \draw (-1.5,0) to[out=30,in=150] (1.5,0);
      \draw (-1.65,0.15) to[out=-50,in=230] (1.65,0.15);
      \draw[->] (-2.3,-1) arc(210:230:2.8 and 2) node[right] {$\phi$};
      \draw[->] (-2.3,-1) arc(160:138:2.8 and 1.8) node[right] {$t$};
    \end{scope}
    \begin{scope}[shift={(8,0)}]
      \draw (0,0) ellipse (3cm and 1.6cm);
      \draw (-2.45,0) to[out=40,in=140] (2.45,0);
      \draw (-2.55,0.1) to[out=-45,in=225] (2.55,0.1);
      \draw (0,-1.6) to[out=180,in=180] (0,-0.95);
      \draw[densely dashed] (0,-1.6) to[out=0,in=0] (0,-0.95);
      \draw ({{cos(140)*3}},{{sin(140)*1.6}}) to[out=30,in=30] ({{cos(145)*2.2}},{{sin(145)*0.8}});
      \draw[densely dashed] ({{cos(140)*3}},{{sin(140)*1.6}}) to[out=210,in=210] ({{cos(145)*2.2}},{{sin(145)*0.8}});
      \draw ({{cos(40)*3}},{{sin(40)*1.6}}) to[out=150,in=150] ({{cos(35)*2.2}},{{sin(35)*0.8}});
      \draw[densely dashed] ({{cos(40)*3}},{{sin(40)*1.6}}) to[out=330,in=330] ({{cos(35)*2.2}},{{sin(35)*0.8}});
      \draw[blue,densely dashed] ({{cos(300)*2.75}},{{sin(300)*1.3}}) arc(-60:120:2.75 and 1.3);
      \fill[color=blue] ({{cos(300)*2.75}},{{sin(300)*1.3}}) circle(0.07);
      \fill[color=blue] ({{cos(120)*2.75}},{{sin(120)*1.3}}) circle(0.07);
    \end{scope}
  \end{tikzpicture}
  \caption{From the CFT perspective, the covering space at finite temperature (right) is a torus whose space direction is $m$ times larger than the original torus (left). Equivalently, the modular parameter is $\tau/m$. On the right, the case $m=3$ is illustrated where the space direction can be split up into $3$ parts, each fully covering the torus on the left. The subsystem on the covering space relevant for entwinement is indicated by a dashed line. In Euclidean signature, the bulk covering space is a solid torus that fills in the boundary covering space (thermal AdS$_3$ or the BTZ black hole with inverse temperature $\beta/m$).}
  \label{fig:BTZ-covering-space}
\end{figure}

We now turn to thermal states. In this case, it is also possible to identify entwinement with an ordinary entanglement entropy in a covering space, however the covering theory differs from that of the conical defect.
As we are looking at a system at finite temperature and finite size, the CFT lives on a torus in Euclidean signature.
The covering theory has central charge $c \to c/m$ and lives on a torus with modular parameter $\tau \to \tau/m$ (see \figref{fig:BTZ-covering-space}).
Its space of states can be read off from the mixed state obtained by projecting the thermal state of the $S_N$ orbifold theory to the subset $\cS_m$ of twisted sectors relevant for entwinement.
The density matrix $\rho_\text{subset}(\tau)$ obtained by this procedure is a mixture of states suppressed by Boltzmann factors $e^{-\beta E}$.
The states in this mixture are contained in the Hilbert space of an $S_{N/m}$ orbifold theory, but not all states of this Hilbert space contribute.
Therefore, the covering theory is not simply the same orbifold theory with a different rank of its gauge group $N \to N/m$ and at a different inverse temperature $\beta \to \beta/m$, but rather an entirely new theory.
We will provide a characterization of this covering theory by decomposing $\rho_\text{subset}(\tau)$ into a sum of Virasoro characters on the covering space, that is characters for a theory with central charge $c/m$ and inverse temperature $\beta/m$.
Each Virasoro character is the contribution from a primary operator on the covering space.
This set of primary operators determines a set of bulk fields on a (bulk) covering space and thus determines the input needed to apply the FLM formula to entwinement.

To begin, we will show that the space of states of the subset of twisted sectors under consideration is a subset of the space of states of an $S_{N/m}$ orbifold theory on the covering space.
Let us first consider the CFT at the orbifold point and postpone the discussion of marginal deformations away from this point for later.
The thermal partition function for the full $S_N$ orbifold theory, which is the starting point of the discussion, is given by the following recursive formula \cite{Gerbershagen:2021gvc}
\begin{equation}
    Z_N(\tau,\tilde Z) = \frac{1}{N} \sum_{k=1}^N \sum_{l=1}^{\lfloor N/k \rfloor} \sum_{j=0}^{k-1} \tilde Z\left(\frac{\tau l+j}{k}\right) Z_{N-kl}(\tau,\tilde Z)
    \label{eq:recursion-relation-partition-function-S_N-orbifold}
\end{equation}
where $\tilde Z(\tau)$ is the seed partition function.
This partition function of course determines the thermal state since we can read off the energy levels and multiplicities from it.
Restricting to the subset of twisted sectors under consideration for entwinement means projecting out some of the terms in the decomposition of the Hilbert space into twisted sectors,
\begin{equation}
  \rho(\tau) \to \rho_\text{subset}(\tau) = \frac{P_m \rho(\tau) P_m}{\Tr(P_m \rho(\tau) P_m)} \quad \text{where} \quad P_m = \sum_{\ket\psi \in \cS_m} \ket\psi\bra\psi
\end{equation}
Here, $\cS_m$ is the subset of twisted sectors containing only long strands of length $km$ ($k \in \ZZ$).
The total Hilbert space of the $S_N$ orbifold theory is given as direct sum of $\cS_m$ and the Hilbert spaces for the remaining twisted sectors.
Since there is a one-to-one correspondence between the thermal state and the partition function, we can describe the effect of the projection $\rho \to \rho_\text{subset}$ by its effect on the partition function.
The projected partition function takes a similar form as \eqref{eq:recursion-relation-partition-function-S_N-orbifold}, however now the symmetric group is replaced by $S_{N/m}$ and the seed partition function is replaced by $Z_{(m)}(\tau) = \frac{1}{m}\sum_{j=0}^{m-1}\tilde Z(\frac{\tau+j}{m})$, the contribution of a long string of length $m$ \cite{Gerbershagen:2021gvc},
\begin{equation}
  Z_{\text{subset},N}^{(m)}(\tau) = Z_{N/m}(\tau,Z_{(m)}) = \frac{m}{N} \sum_{k=1}^{N/m} \sum_{l=1}^{\lfloor N/km \rfloor} \sum_{j=0}^{k-1} Z_{(m)}\left(\frac{\tau l+j}{k}\right) Z_{\text{subset},N-klm}^{(m)}(\tau).
  \label{eq:projected-partition-function-orbifold}
\end{equation}
The contribution of a long string is characterized by a restriction on the allowed spins\footnote{In fact, the same restriction on the allowed spins applies to the conical defect case. The partition function of the $\ZZ_m$ orbifold theory for the conical defect is the same as the contribution \eqref{eq:contribution-long-strand} of a long strand of length $m$ to the $S_N$ orbifold partition function, although the seed theory will differ. Because the ground state of the twisted sector is spinless, the projection plays no role for entwinement in the conical defect case.},
\begin{equation}
    \begin{aligned}
      Z_{(m)}(\tau) &= \frac{1}{m}\sum_{j=0}^{m-1}\tilde Z\left(\frac{\tau+j}{m}\right) = \sum_{(h,\bar{h}) \in \tilde I} q^{\frac{h}{m} - \frac{\tilde c}{24 m}}\bar{q}^{\frac{\bar{h}}{m} - \frac{\tilde c}{24 m}}\frac{1}{m}\sum_{j=0}^{m-1} e^{\frac{2\pi i (h-\bar{h})j}{m}}\\
                 &= \sum_{(h,\bar{h}) \in \tilde I, (h-\bar{h})/m \in \ZZ} q^{\frac{h}{m} - \frac{\tilde c}{24 m}}\bar{q}^{\frac{\bar{h}}{m} - \frac{\tilde c}{24 m}},
    \end{aligned}
    \label{eq:contribution-long-strand}
\end{equation}
where $q=e^{2\pi i \tau}$ and the spectrum of (primary and descendant) conformal dimensions of the seed theory has been denoted by $\tilde I$.
This ensures that the partition function \eqref{eq:projected-partition-function-orbifold} admits only integer spins,
\begin{equation}
  Z^{(m)}_{\text{subset},N}(\tau) = \sum_{(h,\bar{h})} q^{h-\frac{N}{m}\frac{\tilde c}{24}} \bar{q}^{\bar{h}-\frac{N}{m}\frac{\tilde c}{24}} \quad \text{with} \quad h-\bar{h} \in \ZZ.
\end{equation}
Therefore, only states with integer spin appear in $\rho_\text{subset}$,
\begin{equation}
  \rho_\text{subset}(\tau) = \frac{1}{Z^{(m)}_{\text{subset},N}(\tau)} \sum_{(h,\bar{h})} q^{h-\frac{N}{m}\frac{\tilde c}{24}} \bar{q}^{\bar{h}-\frac{N}{m}\frac{\tilde c}{24}}\ket{h,\bar{h}}\bra{h,\bar{h}} \quad \text{with} \quad h-\bar{h} \in \ZZ.
\end{equation}

Now compare this to the thermal state for the $S_{N/m}$ orbifold theory.
Again, we can describe the state by the partition function, which is given by
\begin{equation}
  Z_{N/m}\left(\frac{\tau}{m},\tilde Z\right) = \frac{m}{N} \sum_{k=1}^{N/m} \sum_{l=1}^{\lfloor N/km \rfloor} \sum_{j=0}^{k-1} \tilde Z\left(\frac{\tau}{m}\frac{l}{k}+\frac{j}{k}\right) Z_{\text{covering},N-klm}^{(m)}(\tau).
  \label{eq:partition-function-covering-space}
\end{equation}
To see the difference to the partition function for the subset of twisted sectors described above, consider the $\tilde Z(\tau/m)$ factor.
It differs from the contribution of a long string of length $m$ only in the fact that it lacks the restriction on the allowed spins,
\begin{equation}
  \tilde Z(\tau/m) = \sum_{(h,\bar{h}) \in \tilde I} q^{\frac{h}{m} - \frac{\tilde c}{24 m}}\bar{q}^{\frac{\bar{h}}{m} - \frac{\tilde c}{24 m}}.
\end{equation}
This implies that the spectrum of spins for $Z^{(m)}_{\text{subset},N}$ is a subset of those of $Z_{N/m}\left(\frac{\tau}{m},\tilde Z\right)$: the naive partition function on the covering space allows fractional spins,
\begin{equation}
  Z_{N/m}\left(\frac{\tau}{m},\tilde Z\right) = \sum_{(h,\bar{h})} q^{h-\frac{N}{m}\frac{\tilde c}{24}} \bar{q}^{\bar{h}-\frac{N}{m}\frac{\tilde c}{24}} \quad \text{with} \quad h-\bar{h} \in \ZZ/m.
\end{equation}
It is important to note that the conformal dimensions $(h,\bar{h})$ and thus also the spins we are talking about here are defined w.r.t.~the torus with modular parameter $\tau$, not the one with modular parameter $\tau/m$.
That is, the conformal dimensions are the exponents in an expansion in $q,\bar{q}$ and not in an expansion in $q^{1/m},\bar{q}^{1/m}$.
Of course, in terms of the modular parameter $\tau/m$, spins will always be integer-valued.
But the actual physical modular parameter is $\tau$ while the covering space interpretation is just a convenient computation trick. 

To summarize, $\rho_\text{subset}(\tau)$ appears as a particular state from the perspective of the $S_{N/m}$ orbifold theory which is not purely a thermal state in this theory because it is missing terms with fractional spins in the Boltzmann sum.
Nevertheless, entwinement is equal to an ordinary entanglement entropy for an interval of length $(L+w)/m$ in the particular (non-thermal) state $\rho_\text{subset}(\tau)$.

\medskip
We will now reinterpret the state $\rho_\text{subset}(\tau)$ as arising from the thermal state of a new theory on the covering space, whose spectrum (i.e.~primary operator dimensions and multiplicities) will be determined below.
This will allow us to apply the FLM formula in this new theory.

To determine $1/N$ corrections from the bulk perspective, we need to know the classical background geometry and the quantum fluctuations around the background, i.e.~the set of bulk fields.
The classical background is the same for the thermal state of the $S_{N/m}$ orbifold theory and for the non-thermal state $\rho_\text{subset}(\tau)$.
To see this, note that for low temperatures $\beta > 2\pi m$ (above the Hawking page transition in the covering space) the corresponding partition functions are dominated by the Virasoro character with lowest conformal weight.
This property is called ``vacuum block dominance'' and is a hallmark property of holographic conformal field theories \cite{Hartman:2013mia,Hartman:2014oaa}.\footnote{While not true for any conformal block in $S_N$ orbifolds, see e.g.~\cite{Belin:2017nze} for a counterexample, vacuum block dominance is fulfilled for the genus one partition function. In \secref{sec:monodromy-method}, we will consider conformal blocks for higher genus (replica) partition functions. In that case vacuum block dominance may not hold true universally but will at least be true for a finite region in the $m,w,L$ parameter space, see e.g.~\cite{Gerbershagen:2021yma} for a discussion in the closely related case of ordinary entanglement at finite temperature and finite size.}
In particular, this assumption is reasonable both at the orbifold point as well as far away from it where the CFT is strongly coupled and dual to weakly coupled supergravity.
For $Z_{N/m}(\tau/m,\tilde Z)$, the dominant Virasoro character is simply the vacuum character
\begin{equation}
  Z_{N/m}\left(\frac{\tau}{m},\tilde Z\right) \propto |\chi_0^{(c/m)}(\tau/m)|^2 + O(e^{-c}).
\end{equation}
For the subset partition function, the lowest dimension state is the ground state of the twisted sector containing $N/m$ long strands of length $m$,
\begin{equation}
  Z^{(m)}_{\text{subset},N}(\tau) \propto \left|\chi_{\frac{c}{24}(1-1/m^2)}^{(c)}(\tau)\right|^2 + O(e^{-c})
\end{equation}
From the well-known explicit expressions for the Virasoro characters
\begin{equation}
  \chi_0^{(c)}(\tau) = \frac{q^{-\frac{c-1}{24}}(1-q)}{\eta(\tau)}, \quad \chi_{h>0}^{(c)} = \frac{q^{h-\frac{c-1}{24}}}{\eta(\tau)},
\end{equation}
we see that both partition functions differ only by $O(c^0)=O(N^0)$ terms.
Therefore, in both cases the classical background geometry is thermal AdS$_3$ with temperature $\beta \to \beta/m$ and Newton's constant $G_N \to G_N m$.
For high temperatures, the partition functions are dominated by modular transformed vacuum characters \cite{Gerbershagen:2021gvc},
\begin{equation}
  \begin{aligned}
    Z_{N/m}\left(\frac{\tau}{m},\tilde Z\right) &\propto |\chi_0^{(c/m)}(-m/\tau)|^2 + O(e^{-c}),\\
    Z^{(m)}_{\text{subset},N}(\tau) &\propto |\chi_0^{(c)}(-1/\tau)|^2 + O(e^{-c}).
  \end{aligned}
\end{equation}
Here, the classical background is a BTZ black hole, again with temperature $\beta \to \beta/m$ and Newton's constant $G_N \to G_N m$.

The quantum fluctuations on top of the background, however, depend on which case is considered.
We can reinterpret the restriction on the allowed spins for the subset $\cS_m$ of twisted sectors as changing the set of bulk fields.
To be precise, consider the dominant Virasoro character for the subset partition function $\chi_{\frac{c}{24}(1-1/m^2)}^{(c)}(\tau)$.
We would like to rewrite this character by expanding in characters on the covering space,
\begin{equation}
  \chi_{\frac{c}{24}(1-1/m^2)}^{(c)}(\tau) = \sum_k a_k \chi_{h_k}^{(c/m)}(\tau/m).
\end{equation}
One can think of this expansion as giving us a set of operator dimensions $h_k$ and multiplicity factors $a_k$ that together determine a spectrum of primary operators, although this analogy is somewhat flawed because the ``multiplicities'' $a_k$ can be negative.\footnote{Note that the fact that some of the $a_k$ can be negative does not indicate any inconsistency or non-unitarity of the theory. It simply means that, when interpreted in a covering space picture, some of the terms that would usually come with positive coefficients in the computation of the partition function now must be multiplied with negative ones. But the covering space is merely a computation trick. When interpreted in the actual physical space, all multiplicities are positive.}
A CFT with this primary spectrum has a partition function $Z_\text{covering}(\tau)$ that to the leading and first subleading order in $N$ is equal to the subset partition function,
\begin{equation}
  Z_\text{covering}(\tau) = \sum_{k,k'} a_k a_{k'} \chi_{h_k}^{(c/m)}(\tau/m)\chi_{\bar{h}_{k'}}^{(c/m)}(\bar{\tau}/m) = Z^{(m)}_{\text{subset},N}(\tau).
\end{equation}
Through the standard AdS/CFT dictionary, the primary spectrum determines the set of dual bulk fields.
In addition to these bulk fields from the rewriting of the dominant character with $h=h_\text{min}$, further bulk fields from rewriting characters with conformal dimensions $h=h_\text{min}+O(N^0)$ in terms of characters on the covering space are to be included as well.
These characters have the same leading large $N$ behaviour and thus contribute as well to the large $N$ result for the partition function.
In this way, the ingredients necessary to derive $1/N$ corrections for entwinement can be derived.

We now show how this rewriting of Virasoro characters is accomplished in practice.
Let us first consider low temperatures $\beta > 2\pi m$.
In that case, we can rewrite
\begin{equation}
  \frac{q^{\frac{1}{24}}}{\eta(\tau)} = \frac{1}{\prod_{k=1}^\infty(1-q^k)} = \frac{\prod_{k=0}^\infty\prod_{l=1}^{m-1}(1-q^{k+l/m})}{\prod_{k=1}^\infty(1-q^{k/m})} = \frac{q^{\frac{1}{24m}}}{\eta(\tau/m)}\prod_{k=0}^\infty\prod_{l=1}^{m-1}(1-q^{k+l/m})
\end{equation}
so that
\begin{equation}
  \begin{aligned}
    \chi_{\frac{c}{24}(1-1/m^2)}^{(c)}(\tau) &= \frac{q^{-\frac{c-m}{24m^2}}(1-q^{1/m})}{\eta(\tau/m)} + \frac{q^{-\frac{c-m}{24m^2}}(1-q^{1/m})}{\eta(\tau/m)} \left[-(1-q^{1/m}) + \prod_{k=0}^\infty\prod_{l=1}^{m-1}(1-q^{k+l/m})\right]\\
    &= \chi_0^{(c/m)}(\tau/m) + \sum_{k=0}^\infty\sum_{l=1}^{m-1} a_{k,l} \chi_{km+l}^{(c/m)}(\tau/m).
  \end{aligned}
\end{equation}
The coefficients $a_{k,l}$ are determined by multiplying out the last term in the first line.
Characters of excited states within the subset partition function can be rewritten in the same way.

Similarly, we can express the dominant character at high temperature, $\chi_0^{(c)}(-1/\tau)$, in terms of covering space characters $\chi_h^{(c/m)}(-m/\tau)$.
We first rewrite as follows
\begin{equation}
  \frac{{\tilde q}^{\frac{1}{24}}}{\eta(-1/\tau)} = \sum_{k=0}^\infty p(k) {\tilde q}^k = \sum_{k=0}^\infty\sum_{l=0}^{m-1} p(km+l){\tilde q}^{km+l}
\end{equation}
where $\tilde q = e^{-2\pi i/\tau}$ and $p(n)$ denotes the number of partitions of $n$ into positive integers, $n = \sum_{k=1}^n k n_k$.
We then use the relation
\begin{equation}
    \sum_{k=0}^\infty p(mk+l) {\tilde q}^k = \left(\sum_{k=0}^\infty p(k){\tilde q}^k\right)\left(\sum_{k=0}^\infty p_m(mk+l){\tilde q}^k\right) = \frac{{\tilde q}^{\frac{1}{24}}}{\eta(-1/\tau)}\sum_{k=0}^\infty p_m(mk+l){\tilde q}^k
    \label{eq:relation-for-rewriting-eta}
\end{equation}
where $p_m(n)$ is the number of partitions of $n$ into integers where the same number occurs fewer than $m$ times, i.e.~$n=\sum_{k=1}^n k\, n_k$ where $n_k < m$.
This gives
\begin{equation}
  \frac{{\tilde q}^{\frac{1}{24}}}{\eta(-1/\tau)} = \sum_{l=0}^{m-1}\left(\sum_{k=0}^\infty p(k){\tilde q}^{km+l}\right)\left(\sum_{k=0}^\infty p_m(km+l) {\tilde q}^{km}\right) = \sum_{l=0}^{m-1}\sum_{k=0}^\infty \frac{p_m(km+l){\tilde q}^{km+l+m/24}}{\eta(-m/\tau)}
\end{equation}
We then find
\begin{equation}
  \chi_0^{(c)}(-1/\tau) = \sum_{l=0}^{m-1}\sum_{k=0}^\infty p_m(km+l) \chi_{k+l/m}^{(c/m)}(-m/\tau) + \chi_1^{(c/m)}(-m/\tau) - \chi_{1/m}^{(c/m)}(-m/\tau).
  \label{eq:decomposition-subset-character-covering-character}
\end{equation}
In this expansion, all the multiplicity factors are actually positive.
An expansion of $\chi_h^{(c)}(-1/\tau)$ in terms of $\chi_h^{(c/m)}(-m/\tau)$ is obtained analogously.
Therefore, at high temperatures entwinement is determined by an ordinary entanglement entropy on a covering space given by a BTZ black hole with temperature $\beta/m$ and Newton's constant $G_N m$ where the set of bulk fields is determined by the standard AdS/CFT dictionary from the primary spectrum encoded in \eqref{eq:decomposition-subset-character-covering-character}.
We note again that this computation assumes vacuum block dominance as in \cite{Gerbershagen:2021gvc} but does not require the CFT to be at the orbifold point.

\medskip
We have now shown how to rewrite entwinement as an ordinary entanglement entropy in the covering theory.
Thus, the $1/N$ corrections follow from the FLM formula \cite{Faulkner:2013ana},
\begin{equation}
  S_{w,m}(L) = \frac{\gamma_{L,w}}{4m G_{N,\text{bare}}} + S_\text{bulk,covering} + O(G_{N,\text{bare}}).
\end{equation}
Here, $\gamma_{L,w}$ is the length of a geodesic with opening angle $L$ and winding number $w$ in the BTZ geometry.
This geodesic becomes a Ryu-Takayanagi surface, i.e.~geodesic with opening angle $(L+w)/m$ and zero winding number in the covering space (see \figref{fig:geodesics-BTZ}).
Therefore, we can associate to this geodesic an entanglement wedge in the covering space as well as an entanglement entropy of bulk fields $S_\text{bulk,covering}$ within this entanglement wedge.
Because the set of dual bulk fields for the covering theory differs from that of the original $S_N$ orbifold theory, the $1/N$ correction $S_\text{bulk,covering}$ will differ as well.
But the basic computation method remains applicable.

In fact, the equivalence of entwinement to an ordinary entanglement entropy in the covering theory applies beyond the leading $1/N$ correction.
As shown in app.~\ref{app:covering-space}, the rewriting of the subset partition function as a partition function on the covering space with appropriate spectrum and multiplicity can be done exactly, instead of just for the dominant Virasoro character that we considered above.
Therefore, at least in principle an exact result in $N$ for set of bulk degrees of freedom dual to the subset of twisted sectors considered for entwinement can be obtained.
This implies that the QES prescription \cite{Engelhardt:2014gca} applies to entwinement as well,
\begin{equation}
  S_{w,m}(L) = \min\left\{\frac{\gamma_{L,w}}{4m G_{N,\text{bare}}} + S_\text{bulk,covering}\right\},
\end{equation}
where $S_\text{bulk,covering}$ is the bulk entanglement entropy in the covering space entanglement wedge for the bulk fields determined from the results in app.~\ref{app:covering-space}.

\section{$1/N$ corrections from the monodromy method}
\label{sec:monodromy-method}
While in the previous section, we derived formally how to apply the FLM formula to entwinement, in this section we will be more concretely concerned with the question how large these quantum corrections are in thermal states dual to the BTZ black hole.
To obtain quantitative results, we study quantum corrections from the CFT side using methods from \cite{Barrella:2013wja}.
Our results will be valid for small covering parameter $m \ll N$ and thus small winding number.
Moreover, the results only apply for the small interval phase, where the interval size $(L+w)/m$ in the covering space is not too large such that the leading order result for entwinement is dual to the length of a single winding geodesic.
Large covering interval sizes, where the leading order result for entwinement is dual to the thermal entropy plus the length of a complementary winding geodesic, are considered in app.~\ref{app:covering-space}.
Large winding numbers are studied in the \secref{sec:large-winding-numbers}.

As for the ordinary entanglement entropy, entwinement can be computed by a replica trick,
\begin{equation}
  S_{w,m}(L) = \lim_{n \to 1} \frac{1}{1-n} \log \Tr(\rho_{w,m}(L)^n)
  \label{eq:Renyi-entropy}
\end{equation}
where $n$ is the replica index.
The trace of the associated reduced density matrix is given as a ratio of partition functions,
\begin{equation}
  \Tr(\rho_{w,m}(L)^n) = \frac{Z_n}{Z_1^n}.
  \label{eq:Renyi-entropy-2}
\end{equation}
In \cite{Gerbershagen:2021gvc}, the leading order contribution in $c \propto N$ to this ratio of partition functions was determined by noting that at large $N$ the partition function $Z_n$ is dominated by a single conformal block whose leading order can be obtained from a monodromy method.
To compute $1/N$ corrections, we will compute first the $1/N$ correction to this conformal block using the methods of \cite{Barrella:2013wja}.
The starting point of the method is the so-called decoupling equation associated to the partition function on the higher genus Riemann surface (or equivalently to a two-point function of replica twist operators on the torus) \cite{Gerbershagen:2021yma},
\begin{equation}
  \biggl[
  \partial_z^2 + \sum_{i=1,2} 
  \left(
    \frac{6 h_i}{c}(\wp(z-z_i) + 2\eta_1)
    + \partial_{z_2} f_\text{cl.} (-1)^{i+1}(\zeta(z-z_i) + 2\eta_1 z_i)
  \right)
  - 2\pi i\partial_\tau f_\text{cl.}
  \biggr]\Psi(z) = 0.
  \label{eq:decoupling-equation}
\end{equation}
Here, $z_{1,2}$ are endpoints of the entangling interval along which the $n$ copies of the torus are glued together.
Thus, for entwinement we have $z_1 = 0$ and $z_2 = L + w$.
In the interpretation as a two-point function of twist operators, $h_{1,2}$ are the conformal weights of the replica twist operators which for us are $\frac{6 h_i}{c} = \frac{\Delta}{2m}$ where $\Delta = \frac{1}{2}(1-1/n^2)$.
Note the $1/m$ factor due to the fact that only a $1/m$ fraction of the fields are glued together along the interval $[0,L]$ \cite{Gerbershagen:2021gvc}.
The function $\Psi(z) = \ev{\mathcal{O}_q(z_0)\hat \Psi(z) \mathcal{O}_p(z_1) \mathcal{O}_q(z_\infty)}/\ev{\mathcal{O}_q(z_0) \mathcal{O}_p(z_1) \mathcal{O}_q(z_\infty)}$ is a ratio of correlation functions where $\hat \Psi(z)$ is a conformal primary operator of weight $h_\Psi = -1/2 + O(1/c)$ (see \cite{Gerbershagen:2021yma} for details).
Finally, imposing certain monodromy conditions on the solution of the decoupling equation determines the semiclassical conformal block $f_\text{cl.}$.
Depending on the choice of monodromy conditions, conformal blocks in different channels can be obtained.

The first subleading order in the $1/N$ expansion of the conformal block is determined from the geometry of the higher genus Riemann surface in question.
Any compact Riemann surface of genus $g$ can be parametrized as a quotient of the complex plane by a Schottky group $G_S$, that is a subgroup of $\PSLtwoC$ freely generated by $g$ $\PSLtwoC$ elements $L_1,...,L_g$.
Let us denote Schottky group elements by $\Gamma=\prod_k L_{i_k}^{n_k}$ and their eigenvalues by $Q_\Gamma^{\pm1/2}$.
A Schottky group element $\Gamma$ is called primitive if it cannot be written as a power $\tilde \Gamma^n$ for any $\tilde\Gamma \in G_S$ and $n>1$.
Let us denote the set of primitive conjugacy classes of $G_S$ by $P(G_S)$.
The conformal block then acquires subleading $1/N$ corrections depending on the set of bulk fields,
\begin{equation}
  \left.\log(Z_n)\right|_{O(N^0)} = \delta \log Z_{n,\text{metric}} + \sum_i \delta \log Z_{n,\text{scalar}}(m_i) + ...
\end{equation}
where for simplicity we have only written down contributions due to metric and scalar field fluctuations.
$m_i$ is the mass of the $i$-th bulk scalar field.
Explicitly, the $1/N$ corrections are given by \cite{Yin:2007gv,Giombi:2008vd,Barrella:2013wja}
\begin{equation}
  \begin{aligned}
    \delta \log Z_{n,\text{metric}} &= -2\sum_{\Gamma \in P_S}\sum_{k=2}^\infty \log|1-Q_\Gamma^k|,\\
    \delta \log Z_{n,\text{scalar}}(m_i) &= -\sum_{\Gamma \in P_S}\sum_{l,l'=0}^\infty \log|1-Q_\Gamma^{l+\Delta_i/2}\bar{Q}_\Gamma^{l'+\Delta_i/2}|,
  \end{aligned}
  \label{eq:1-over-N-correction-partition-function}
\end{equation}
where $m_i^2 = \Delta_i(\Delta_i-2)$.
The spectrum $\Delta_i$ is the one of the covering theory determined in the previous section.

Therefore, the task is now to determine the Schotty generators $L_1,...,L_g$ which will then determine the Schottky group elements and their eigenvalues.
Fortunately, the same decoupling equation that allows us to determine the leading semiclassical conformal block also allows us to determine the Schottky generators.
The decoupling equation also determines a uniformization mapping from the branched cover of $n$ copies of the torus to the complex plane on which the Schotty group acts.
If $z$ denotes the coordinate on the branched cover and $\tilde z$ denotes the coordinate on the complex plane, then this uniformization map is given by the ratio of two independent solutions of the decoupling equation, $\tilde z=\Psi_+(z)/\Psi_-(z)$.
Moving around a closed loop on the Riemann surface induces a monodromy for this pair of solutions,
\begin{equation}
  \begin{pmatrix}
    \Psi_+(z)\\
    \Psi_-(z)
  \end{pmatrix}
  \to
  \begin{pmatrix}
    a & b\\
    c & d
  \end{pmatrix}
  \begin{pmatrix}
    \Psi_+(z)\\
    \Psi_-(z)
  \end{pmatrix}.
\end{equation}
This monodromy gives a $\PSLtwoC$ action in $\tilde z$ coordinates,
\begin{equation}
  \tilde z \to \frac{a\tilde z+b}{c\tilde z+d}.
\end{equation}
We have fixed the monodromy around two of the closed loops on the Riemann surface\footnote{We assume $\ZZ_n$ replica symmetry, which allows us to consider only two of the loops \cite{Gerbershagen:2021yma}.} in order to determine the semiclassical conformal block from its derivates $\partial_\tau f_\text{cl.},\partial_{z_2} f_\text{cl.}$.
From the monodromy for a basis of the remaining loops we obtain the Schottky generators, $L_i = \begin{pmatrix} a & b\\ c & d \end{pmatrix}$.

For the leading order entwinement result, it is sufficient to solve \eqref{eq:decoupling-equation} in a series expansion around $n-1$.
For the subleading orders, we will need a solution which is valid for arbitrary $n$.
This can be obtained by expanding around high or low temperatures.
For us, the more interesting case is at high temperatures where the dual spacetime is a BTZ black hole hence we start with this case.

In the first step, we change variables to
\begin{equation}
  u = e^{-2\pi i z/\tau}, \quad \tilde q = e^{-2\pi i/\tau}
\end{equation}
since we want to expand in a power series in $\tilde q$.
In these variables, the torus is described by the identifications $u \sim u/\tilde q$.
Moreover, the equations simplify if we also change to a different function
\begin{equation}
  \Psi(z) = \left(\frac{\partial u}{\partial z}\right)^{h_\Psi}\tilde\Psi(u(z))
\end{equation}
and solve the equation in terms of $\tilde\Psi(u)$.
Of course, this is just the transformation rule for how $\Psi(z)$, given as a ratio of correlation functions, transforms under the conformal transformation $z \to u(z)$.
Similary, we will define a new function $\tilde f_\text{cl.}$ by using that we are computing a dominant conformal block and applying the known transformation properties of correlation functions under the conformal transformation $z \to u(z)$,
\begin{equation}
  f_\text{cl.} = \tilde f_\text{cl.} - \sum_{i=1,2} h_i \log\left(\frac{\partial u(z_i)}{\partial z_i}\right).
\end{equation}
We will then make use of the series expansion of the Weierstraß functions \cite{NIST:DLMF}
\begin{equation}
  \begin{aligned}
    &\wp(z) = -\frac{2\eta_3}{\tau} + \left(\frac{2\pi i}{\tau}\right)^2 \sum_{k=-\infty}^\infty \frac{{\tilde q}^k u}{(u - {\tilde q}^k)^2}\\
    &\zeta(z) = \frac{2\eta_3 z}{\tau} -\frac{i\pi}{\tau} \sum_{k=-\infty}^\infty \frac{{\tilde q}^k + u}{{\tilde q}^k - u},
  \end{aligned}
\end{equation}
where $\eta_3 = \tau\eta_1 - \pi i$.
Putting everything together, we find to leading order in the large central charge limit
\begin{equation}
  \tilde\Psi''(u) + \biggl[
  \begin{aligned}[t]
    &\frac{\Delta}{2m}\sum_k \frac{{\tilde q}^k}{u(u-{\tilde q}^k)^2} + \frac{\Delta}{2m} \sum_k \frac{{\tilde q}^k u_2}{u(u-{\tilde q}^k u_2)^2}\\
    &+ \left(\frac{\Delta}{2m} - u_2 \partial_{u_2} \tilde f_\text{cl.}\right) \sum_k \frac{{\tilde q}^k(u_2-1)}{u(u-{\tilde q}^k)(u-{\tilde q}^k u_2)} + \frac{1/4 - {\tilde q} \partial_{\tilde q} \tilde f_\text{cl.}}{u^2}
  \biggr]\tilde\Psi(u) = 0
  \end{aligned}
\end{equation}
where $u_2 = u(z_2) = u(L) {\tilde q}^{w} = u_L {\tilde q}^{w}$.

For small winding number $w$, it is known that the monodromy condition that gives the dominant conformal block is trivial monodromy around a) the time circle and b) the entangling interval plus $w$ times the spatial circle.
For this monodromy, the high temperature limit we need corresponds to $\tau \to 0 \Leftrightarrow {\tilde q} \to 0$ while keeping $u_2$ fixed.
As $u_2$ depends on ${\tilde q}$, this means that we analytically continue $w$ from the integers to a real number.
This real number scales with ${\tilde q}$ such that $u_2$ stays constant as ${\tilde q}$ goes to zero.
Only in the end we set $w$ to be an integer again.
As a consistency check, we note that this procedure gives the same result as the expansion in $n-1$ used in \cite{Gerbershagen:2021gvc}.
We find that in the limit $\tilde q \to 0$, keeping $u_2$ fixed, the decoupling equation becomes
\begin{equation}
  \begin{aligned}
    \tilde\Psi_0''(u) + \biggl[&\frac{\Delta}{2m}\frac{1}{u(u-1)^2} + \frac{\Delta}{2m}\frac{u_2}{u(u-u_2)^2}\\
    +& \left(\frac{\Delta}{2m} - u_2 \partial_{u_2} \tilde f_\text{cl.}\right)\frac{u_2-1}{u(u-1)(u-u_2)} + \frac{1/4 - {\tilde q}\partial_{\tilde q} \tilde f_\text{cl.}}{u^2}\biggr]\tilde\Psi_0(u) = 0.
  \end{aligned}
\end{equation}
At $\tau=0$, the torus degenerates into a cylinder.
In $u$ coordinates, a fundamental domain for the torus is given by an annulus in the complex plane that becomes thicker and thicker as ${\tilde q} \to 0$ until at ${\tilde q}=0$ it covers the entire $u$ plane.
Therefore, we can apply the known formula for the two-point function on the plane in order to compute $\tilde f_\text{cl.}$ at ${\tilde q}=0$,
\begin{equation}
  \ev{\mathcal{O}(1) \mathcal{O}(u_2)} = (u_2-1)^{-2 h} = e^{-\frac{c}{6}\tilde f_\text{cl.}}/Z(\tau)
\end{equation}
where $Z(\tau)$ is the thermal partition function which in the ${\tilde q} \to 0$ limit goes as $Z(\tau) \sim {\tilde q}^{-c/24}$.
This gives
\begin{equation}
  \tilde f_\text{cl.} = \frac{\Delta}{m} \log(u_2-1) + \frac{1}{4}\log {\tilde q}.
  \label{eq:semiclassical-conformal-block-q=0}
\end{equation}

Let us quickly check that this is consistent with the known result for entwinement in this limit, which is quoted in \cite{Gerbershagen:2021gvc} as
\begin{equation}
  S_{w,m}(L) = \frac{c}{3m} \log\biggl[\frac{\beta}{2\pi^2\epsilon_\text{UV}} \sinh\bigl(\frac{2\pi^2(L+w)}{\beta}\bigr)\biggr].
  \label{eq:entwinement-result-high-temperature}
\end{equation}
Transforming back to $f_\text{cl.}$, we find from \eqref{eq:semiclassical-conformal-block-q=0}
\begin{equation}
  \begin{aligned}
    f_\text{cl.} &= \tilde f_\text{cl.} - \frac{\Delta}{m}\log\left(\frac{2\pi i}{\tau}\right) - \frac{\Delta}{2m}\log(u_2)\\
                 &= \frac{\Delta}{m} \log\left(\frac{\tau}{\pi i}\sinh\left(\frac{\pi i}{\tau}(L+w)\right)\right) - \frac{\pi i}{2\tau}
  \end{aligned}
\end{equation}
in agreement with \eqref{eq:entwinement-result-high-temperature} for $\beta = -2\pi i \tau$ and $Z_n = e^{-\frac{c}{6} f_\text{cl.}}$.

Coming back to solving the decoupling equation, at ${\tilde q}=0$ we are left with
\begin{equation}
  \tilde\Psi_0''(u) + \frac{\Delta}{2m}\frac{(u_2-1)^2}{(u-1)^2(u-u_2)^2}\tilde\Psi_0(u) = 0.
\end{equation}
The solution to this equation is given by
\begin{equation}
  \tilde \Psi_{0,\pm}(u) = (u-1)^{\frac{1}{2}(1 \pm \sqrt{1-2\Delta/m})}(u-u_2)^{\frac{1}{2}(1 \mp \sqrt{1-2\Delta/m})}.
  \label{eq:0-th-order-solution-decoupling-equation-high-temperature}
\end{equation}
The higher orders are determined by expanding
\begin{equation}
  \tilde \Psi(u) = \sum_{k=0}^\infty \tilde \Psi_k(u) u^k, \quad \tilde f_\text{cl.} = \frac{\Delta}{m} \log(u_2-1) + \frac{1}{4}\log {\tilde q} + \sum_{k=1}^\infty \tilde f_k {\tilde q}^k.
\end{equation}
For instance, to first order we find
\begin{equation}
  \tilde \Psi_1''(u) + \frac{\Delta}{2m}\frac{(u_2-1)^2}{(u-1)^2(u-u_2)^2}\tilde\Psi_1(u) + \tilde T_1(u) \tilde \Psi_0(u) = 0.
\end{equation}
with
\begin{equation}
  T_1(u) = -\frac{\tilde f_1}{u^2} - \frac{(u_2-1)u_2\partial_{u_2}\tilde f_1}{u(u-1)(u-u_2)}.
\end{equation}
The first order solution for $\tilde\Psi$ is given by
\begin{equation}
  \begin{aligned}
    \tilde\Psi_1(u) = &\frac{\tilde\Psi_{0,-}(u)}{(u_2-1)\sqrt{1-2\Delta/m}}\int du \tilde\Psi_{0,+}(u) T_1(u) \tilde\Psi_0(u)\\
    - &\frac{\tilde\Psi_{0,+}(u)}{(u_2-1)\sqrt{1-2\Delta/m}}\int du \tilde\Psi_{0,-}(u) T_1(u) \tilde\Psi_0(u).
  \end{aligned}
\end{equation}
From trivial monodromy around $u=0$, equivalent to trivial monodromy around the time circle of the torus, we find $\tilde f_1 = 0$.
The solution of the decoupling equation to higher orders in ${\tilde q}$ can be found similarly.\footnote{In practice, it is simpler to also expand the solution in $u$, similar to the treatment in \cite{Barrella:2013wja}, to avoid having to deal with hypergeometric functions in the solution of the decoupling equation.}

To find the $n$ Schottky generators, we need to determine the monodromy around $n$ independent closed paths on the Riemann surface.
Here, the difference between the ordinary entanglement entropy and entwinement comes into play again.
For the ordinary entanglement entropy, one of the closed paths simply goes around the space circle of the torus (remember that we already fixed the monodromy around the time circle),
\begin{equation}
  \begin{pmatrix}
    \tilde \Psi_+(u/{\tilde q})\\
    \tilde \Psi_-(u/{\tilde q})
  \end{pmatrix}
  = M_{\tilde q}
  \begin{pmatrix}
    \tilde \Psi_+(u)\\
    \tilde \Psi_-(u)
  \end{pmatrix}.
\end{equation}
Other closed paths are obtained by combining a path around the space circle with encircling the endpoints of the entangling interval an integer number of times.
For example, starting from the first torus copy we can encircle an interval endpoint clockwise, which lands us on the second copy, then move once around the spatial circle and encircle the endpoint counter clockwise to get back to the same point on the first copy.
As the monodromy of the solutions around these endpoints is known from \eqref{eq:0-th-order-solution-decoupling-equation-high-temperature} we find the Schottky generators
\begin{equation}
  L^\text{EE}_k = M_1^{k-1} M_{\tilde q} M_2^{k-1}
  \quad \text{where} \quad
  M_1 = M_2^{-1} =
  \begin{pmatrix}
    e^{\pi i(1 + \sqrt{1-2\Delta/m})} & 0\\
    0 & e^{\pi i(1 - \sqrt{1-2\Delta/m})}
  \end{pmatrix}.
\end{equation}
For entwinement, we have to keep in mind that the projection onto the subset of twisted sectors implies that the space direction is effectively $m$ times as long.
Thus, going once around the space direction is not a closed loop on the Riemann surface and to find the Schottky generators we have to go around the space direction $m$ times.
The Schottky generators relevant for computing entwinement are then given by
\begin{equation}
  L_k = M_1^{k-1} M_{\tilde q}^m M_2^{k-1}.
  \label{eq:Schottky-generators-entwinement}
\end{equation}
Explicitly, we find for the components of the first Schottky generator
\begin{equation}
  \begin{aligned}
    (L_1)_{11} &= \left.(L_1)_{22}\right|_{\nu \to - \nu} = \frac{u_2^{\frac{1}{2}(1-\nu)}}{\nu(1-u_2)\sqrt{{\tilde q}}}\left(1 - \frac{{\tilde q}}{4u_2}(u_2 + 1 + \nu(u_2-1))^2 + O({\tilde q}^2)\right)\\
    (L_1)_{12} &= -(L_1)_{21} = \frac{\sqrt{u_2}}{\nu(u_2-1)\sqrt{{\tilde q}}}\left(1 - \frac{{\tilde q}}{2 m u_2}(2 m u_2 + \Delta(u_2-1)^2) + O({\tilde q}^2)\right)
  \end{aligned}
\end{equation}
where $\nu = \sqrt{1-2\Delta/m}$.

Moving on the eigenvalues $Q_{L_k}^{\pm1/2}$ of the Schottky generators, we note that $Q_{L_k} = Q_{M_q}^m$ since the $M_1^{k-1},M_2^{k-1}$ parts in \eqref{eq:Schottky-generators-entwinement} cancel in the computation of the eigenvalues and the eigenvalue of the $m$-th power of a matrix is the $m$-th power of the eigenvalue.
To first order in ${\tilde q}$ we get
\begin{equation}
  Q_{M_q} = {\tilde q}\frac{(1-2\Delta/m)u_2^{-1+\sqrt{1-2\Delta/m}}(u_2-1)^2}{(u_2^{\sqrt{1-2\Delta/m}}-1)^2} + O({\tilde q}^2).
  \label{eq:eigenvalue-single-letter-word}
\end{equation}
Inserting this into \eqref{eq:1-over-N-correction-partition-function}, \eqref{eq:Renyi-entropy-2} and \eqref{eq:Renyi-entropy} we obtain the leading $1/N$ correction to entwinement due bulk metric fluctuations
\begin{equation}
  \begin{aligned}
    S_{w,m}(L) &= \frac{c}{3m} \log\biggl[\frac{\beta}{2\pi^2\epsilon_\text{UV}} \sinh\bigl(\frac{2\pi^2(L+w)}{\beta}\bigr)\biggr]\\
               &+ {\tilde q}^{2m}\left[8 - \frac{16\pi^2(L+w)}{\beta}\coth\left(\frac{2\pi^2(L+w)}{\beta}\right)\right] + O({\tilde q}^{4m})\\
               &+ O(1/c).
  \end{aligned}
  \label{eq:1-over-N-corrections-small-intervals}
\end{equation}
$1/N$ corrections due to scalar fields are obtained similarly,
\begin{equation}
    \delta S_{w,m}(L) = -2\Delta_i \tilde q^{m\Delta_i}\left[1-\frac{2\pi^2(L+w)}{\beta}\coth\left(\frac{2\pi^2(L+w)}{\beta}\right)\right] + O({\tilde q}^{2m\Delta_i}).
\end{equation}

Higher order terms in the expansion in ${\tilde q}$ are determined from multi-letter words $\Gamma=L_{k_1}...L_{k_n}$.
In order to derive these contributions, note first that
\begin{equation}
    M_{\tilde q}^m = \left(\frac{\sqrt{u_2}}{\nu\sqrt{{\tilde q}}(1-u_2)}\right)^m (u_2^{-\nu/2}-u_2^{\nu/2})^{m-1} \begin{pmatrix}
    u_2^{-\nu/2} & 1\\
    -1 & -u_2^{\nu/2}
    \end{pmatrix}
    + O({\tilde q}^{-m/2+1}).
\end{equation}
The eigenvalue of a double letter word $\Gamma=L_{k_1}L_{k_2}$ is thus given by
\begin{equation}
    Q_{\Gamma} = {\tilde q}^{2m}\frac{\left(\nu(u_2^{1/2}-u_2^{-1/2})\right)^{4m}}{4\left(u_2^{-\nu/2}-u_2^{\nu/2}\right)^{4(m-1)}\left(2\sin^2\left(\frac{\pi(k_1-k_2)}{n}\right)-1+\frac{u_2^{-\nu}+u_2^\nu}{2}\right)^2}.
\end{equation}
In order to determine the contribution to entwinement from these Schottky group elements, we have to sum $Q_\Gamma^2$ over $k_1,k_2$ from $0$ to $n-1$ and then analytically continue to $n \to 1$.
The only $k_1,k_2$ dependence in $Q_\Gamma$ is in the denominator $2\sin^2\left(\frac{\pi(k_1-k_2)}{n}\right)-x$ where $x=1-\frac{u_2^{-\nu}+u_2^\nu}{2}$.
Therefore, we have to determine the analytic continuation of
\begin{equation}
    \sum_{k=0}^{n-1} \frac{1}{\left(2\sin^2\left(\frac{\pi k}{n}\right)-x\right)^{2p}} = x^{-2p} - \frac{1}{(2p-1)!}\partial_x^{2p-1}\sum_{k=1}^{n-1}\sum_{r=0}^\infty \frac{x^r}{\left(2\sin^2\left(\frac{\pi k}{n}\right)\right)^{r+1}}.
\end{equation}
Using \cite{Calabrese:2010he}
\begin{equation}
    \frac{n}{2}\sum_{k=1}^{n-1} \frac{1}{\left(\sin^2\left(\frac{\pi k}{n}\right)\right)^\alpha} = (n-1)\frac{\sqrt{\pi}}{2}\frac{\Gamma(\alpha+1)}{\Gamma(\alpha+3/2)} + O((n-1)^2)
\end{equation}
where $\Gamma(z)$ is the Gamma function, we get
\begin{equation}
    \sum_{k=0}^{n-1} \frac{1}{\left(2\sin^2\left(\frac{\pi k}{n}\right)-x\right)^{2p}} = x^{-2p} - (n-1)\frac{1}{(2p-1)!}\partial_x^{2p-1}\left(\frac{\arcsin\sqrt{x/2}}{\sqrt{1-x/2}(x/2)^{3/2}} - \frac{2}{x}\right) + O((n-1)^2).
\end{equation}
The contribution to the $1/N$ correction to entwinement from bulk metric fluctuations via double letter words is then
\begin{equation}
    \begin{aligned}
    \Delta S_{w,m}(L) =\, &\frac{{\tilde q}^{4m}}{3\cosh^7\left(2\pi^2(L+w)/\beta\right)}\biggl[30\cosh\left(2\pi^2(L+w)/\beta\right)-106\cosh^3\left(2\pi^2(L+w)/\beta\right)\\
    &+130\cosh^5\left(2\pi^2(L+w)/\beta\right)-42\cosh^7\left(2\pi^2(L+w)/\beta\right)\\
    &-8\cosh^9\left(2\pi^2(L+w)/\beta\right)+8\cosh^{11}\left(2\pi^2(L+w)/\beta\right)\\
    &+\frac{12\pi^2(L+w)}{\beta}\left(-5+6\cosh^2\left(2\pi^2(L+w)/\beta\right)\right)\sinh^5\left(2\pi^2(L+w)/\beta\right)\biggr].
    \end{aligned}
\end{equation}

\section{Finite size corrections for large winding numbers}
\label{sec:large-winding-numbers}
Because the expansion parameter in the large $N$ limit for entwinement is $N/m$ instead of $N$, the results of the preceding section apply to small values of the parameter $m$ and hence small winding numbers $w$.
Here, we discuss the opposite limit of large $m$, more specifically the largest possible value $m=N$.
Following the method from \cite{Cardy:2014jwa}, we determine universal finite size or finite temperature corrections for entwinement.
What we mean by that are corrections of the form
\begin{equation}
  \delta S_{w,m}(L) = S_{w,m}(L) - \left[S_{w,m}(L)\right]_{c \to \infty},
\end{equation}
which determine how much entwinement differs from the length of a bulk geodesic with winding number $w$.
These are the same corrections as computed in the previous section, but since we now consider $m=N \propto c$ the corrections are no longer parametrically suppressed by $1/N$ and thus it is not appropriate to call them $1/N$ corrections.
The term finite size/temperature correction is chosen because the large central charge result takes the same form as the entanglement entropy for an infinitely large system at finite temperature (at high temperature) or a finite size system at zero temperature (at low temperature).
These corrections are of interest for better understanding the ``entanglement builds geometry'' proposal.
They determine how closely entwinement approximates the length of a bulk geodesic even in a parameter regime where the physics in the bulk region probed by the putative dual geodesic is badly approximated by classical GR.
Therefore, the corrections determine if entwinement looks geometrically, in the sense that its value is well approximated by a geodesic length, even though the notion of a classical geometry in which the geodesic is embedded is not a good description of the bulk physics.

The corrections we would like to compute are valid for $S_N$ orbifolds with a gapped seed theory.
They are proportional to the conformal dimension $\Delta_1$ of the lowest weight non-vacuum state.
As we are considering a subset $\cS_m$ of twisted sectors for entwinement and thus also a subset of states, we need to look for the lowest weight conformal dimension within this subset.
For $m=N$ only the maximally twisted sector survives, whose partition function is given by
\begin{equation}
  Z_{(m)}(\tau) = \frac{1}{m} \sum_{j=0}^{m-1} \tilde Z\left(\frac{\tau+j}{m}\right).
\end{equation}

To determine the finite temperature corrections, we expand at low temperatures as follows,
\begin{equation}
  \begin{aligned}
    Z_{(m)}(\tau) &= \frac{1}{m} \sum_{j=0}^{m-1} \sum_{(h,\bar h)} e^{2\pi i (h-\tilde c/24)\frac{\tau+j}{m}} e^{-2\pi i (\bar h-\tilde c/24)\frac{\bar \tau+j}{m}}\\
               &= e^{\frac{\tilde c}{12} \beta m} \sum_{(\Delta,s)} e^{-\beta \left(\frac{\Delta}{m}+\frac{\tilde c}{12}\left(m-\frac{1}{m}\right)\right)} \delta_{s \in m\ZZ}.
  \end{aligned}
\end{equation}
For generic $m$, the lowest weight non-vacuum contribution to this will come from a zero spin state (otherwise the seed theory has to depend on $m$).
The mixed state on the subset of twisted sectors can thus be expanded as
\begin{equation}
  \rho = \frac{\ket{0}\bra{0}+e^{-\Delta_1\beta/m}\ket{\psi_1}\bra{\psi_1}+...}{1+e^{-\Delta_1\beta/m}+...},
\end{equation}
where $\ket{0}$ denotes the seed theory ground state.
The computation of the correction to entwinement proceeds in the same way as in \cite{Cardy:2014jwa}.
By the state-operator correspondence the correction to the Rényi entropy can be related to a two-point function of the operator $\psi_1(z)$ dual to $\ket{\psi_1}$ on a multisheeted Riemann surface given by $n$ copies of the complex plane sewn together along the entangling interval.
The genus of this Riemann surface is zero and thus this two-point function can be computed by applying a uniformization transformation to the complex plane.
This results in
\begin{equation}
  \begin{aligned}
    S_{w,m}(L) = &\frac{c}{3m}\log\left(\frac{1}{\pi\epsilon_\text{UV}}\sin\left(\frac{\pi(L+w)}{m}\right)\right)\\
                 &+ 2g\Delta_1\left(1-\frac{\pi (L+w)}{m}\cot\left(\frac{\pi(L+w)}{m}\right)\right) e^{-\Delta_1\beta/m} + ...
  \end{aligned}
\end{equation}

For the finite size correction, we perform a modular transformation and then expand at high temperatures.
Consider first the case that $m$ is a prime number.
In that case,
\begin{equation}
  \begin{aligned}
    Z_{(m)}(\tau) &= \frac{1}{m} \sum_{j=0}^{m-1} \tilde Z\left(\frac{\tau+j}{m}\right) = \frac{1}{m}\left(\tilde Z\left(-\frac{m}{\tau}\right) + \sum_{j=1}^{m-1} \tilde Z\left(-\frac{1}{m\tau} + \frac{j}{m}\right)\right)\\
               &= \frac{1}{m} e^{\frac{\tilde c}{12}\, \frac{4\pi^2}{\beta}m} \sum_{(\Delta,s)} \left(e^{-\frac{4\pi^2}{\beta} m \Delta} + e^{-\frac{4\pi^2}{\beta}\left(\frac{\Delta}{m} + \frac{\tilde c}{12}\left(m-\frac{1}{m}\right)\right)}(m\delta_{s \in m \ZZ} - 1)\right).
  \end{aligned}
  \label{eq:partition-function-maximally-twisted-sector}
\end{equation}
The leading contribution to the partition function for the maximally twisted sector depends on the gap $\Delta_1$ in the seed theory.
If $\Delta_1 > \frac{\tilde c}{12}$ \footnote{This is not true for instance for the D1/D5 CFT whose seed theory contains 4 free fermions and bosons. On the other hand, a single free fermion does obey this property and for a free compact boson it depends on the compactification radius.}, the leading contribution at $\beta \to 0$ comes from the second term in the last line of \eqref{eq:partition-function-maximally-twisted-sector} and the result for entwinement is given by
\begin{equation}
  S_{w,m}(L) = \frac{c}{3m}\log\left(\frac{\beta}{2\pi^2\epsilon_\text{UV}}\sinh\left(\frac{2\pi^2(L+w)}{\beta}\right)\right) + \delta S_{w,m}(L)
\end{equation}
where
\begin{equation}
  \delta S_{w,m}(L) = 2g\Delta_1\left(1-\frac{1}{m}\right)\left(1-\frac{2\pi^2 (L+w)}{\beta}\coth\left(\frac{2\pi^2(L+w)}{\beta}\right)\right) e^{-\frac{4\pi^2}{\beta}\left(\frac{\Delta_1}{m} + \frac{\tilde c}{12}\left(m-\frac{1}{m}\right)\right)} + ...
  \label{eq:entwinement-maximally-twisted-sector-large-gap}
\end{equation}
On the other hand, for $\Delta_1 < \frac{\tilde c}{12}$, the first term in the last line of \eqref{eq:partition-function-maximally-twisted-sector} gives the leading finite size correction equal to
\begin{equation}
  \delta S_{w,m}(L) = 2g\Delta_1\frac{1}{m}\left(1-\frac{2\pi^2 (L+w)}{\beta}\coth\left(\frac{2\pi^2(L+w)}{\beta}\right)\right) e^{-\frac{4\pi^2}{\beta}m\Delta_1} + ...
  \label{eq:entwinement-maximally-twisted-sector-small-gap}
\end{equation}
When $m$ is not prime, the modular transformation will give different terms.
For instance, for $m=4$,
\begin{equation}
  Z_{(m)}(\tau) = \frac{1}{4}\left(\tilde Z\left(-\frac{4}{\tau}\right) + \tilde Z\left(-\frac{1}{4\tau} + \frac{3}{4}\right) + \tilde Z\left(-\frac{1}{\tau} + \frac{1}{2}\right) + \tilde Z\left(-\frac{1}{4\tau} + \frac{1}{4}\right)\right).
\end{equation}
In general, there are terms of the form $\tilde Z\left(\frac{m}{j^2}\left(-\frac{1}{\tau}+k\right)\right)$ where $j$ divides $m$.
However, in the $\beta \to 0$ limit the leading contributions comes either from $\tilde Z\left(-\frac{m}{\tau}\right)$ for $\Delta_1 < \frac{\tilde c}{12}$ or from $\tilde Z\left(-\frac{1}{m\tau} + \frac{k}{m}\right)$ for $\Delta_1 > \frac{\tilde c}{12}$.
The other terms scale as $e^{-\frac{4\pi^2}{\beta}\frac{m}{j^2}\left(\Delta_1-\frac{\tilde c}{12}\right)}$ for $\beta \to 0$ such that the leading contribution comes either from $j$ being as small as possible ($j=1$) for $\Delta_1 < \frac{\tilde c}{12}$ or from $j$ being as large as possible ($j=m$) for $\Delta_1 > \frac{\tilde c}{12}$.
Therefore, for $m$ not a prime number, entwinement for the maximally twisted sector behaves as follows.
For $\Delta_1 > \frac{\tilde c}{12}$, the result is again given by \eqref{eq:entwinement-maximally-twisted-sector-small-gap}.
For $\Delta_1 < \frac{\tilde c}{12}$ only the prefactor $\left(1-\frac{1}{m}\right)$ in \eqref{eq:entwinement-maximally-twisted-sector-large-gap} changes,
\begin{equation}
  \delta S_{w,m}(L) = 2g\Delta_1\frac{\phi(m)}{m}\left(1-\frac{2\pi^2 (L+w)}{\beta}\coth\left(\frac{2\pi^2(L+w)}{\beta}\right)\right) e^{-\frac{4\pi^2}{\beta}\left(\frac{\Delta_1}{m} + \frac{\tilde c}{12}\left(m-\frac{1}{m}\right)\right)} + ...
\end{equation}
where $\phi(m)$ is the Euler totient function.

In summary, we find that the leading finite size/temperature corrections to entwinement for the maximally twisted sector have the same dependence on the entangling interval size $L$ and winding number $w$ as for small values of $m$.
Only the temperature dependence differs and depends on the gap between vacuum and first excited spinless state in the seed theory.
This implies that in all cases the size of the corrections increases for larger winding numbers.
On the other hand, changes of the parameter $m$ and thus of the central charge lead to an exponential suppression in the magnitude of the finite size corrections.

\section{Discussion and conclusion}
\label{sec:discussion}
This publication contains two new main results.
Firstly, we explained how $1/N$ corrections to entwinement are given formally by applying the well-known FLM \cite{Faulkner:2013ana} or QES \cite{Engelhardt:2014gca} formulas in a fictitious covering space.
This space covers the bulk spacetime dual to the state in question an integer number of times.
The bulk field content on the covering space depends also on the state and on the choice of covering parameter $m$.
While this result is arguably obvious for the states dual to conical defects, its applicability to thermal states is non-trivial.
In particular, the bulk field content in the covering space for a BTZ black hole is related in an intricate, more complicated way to that of the original BTZ black hole than for the conical defect case.
Secondly, we provided explicit results for $1/N$ corrections at finite temperature, focussing on universal results valid for any holographic CFT with an $S_N$ orbifold structure.
We find that for fixed covering parameter $m$, the corrections increase with increasing winding number $w$ while for fixed $w$, there is an exponential suppression with increasing $m$.

\medskip

Our findings have implications for the proposal that the bulk geometry should be encoded in terms of boundary entanglement, summarized under the slogan ``entanglement builds geometry'' \cite{Swingle:2009bg,VanRaamsdonk:2010pw}.
At leading order in large $N$, entwinement is dual to the length of a winding geodesic which can probe deep into the bulk interior, close to singularities or black hole horizons.
Certainly close to a singularity, one expects the notion of a classical spacetime to not be a good description of the physics at hand.
A goal of the present publication is to determine if such a breakdown is visible in the entanglement structure, i.e.~if the approximation of the entanglement entropy by the length of a geodesic gets worse and worse as we approach a horizon or singularity.
For ordinary entanglement entropy, this question is immaterial since the entanglement shadow shields the regions close to horizons or singularities from being probed by Ryu-Takayanagi surfaces.
But this is not the case for entwinement.

Using our results of \secref{sec:monodromy-method} and \ref{sec:large-winding-numbers}, we can draw the following conclusions pertaining to ``entanglement build geometry'' in the BTZ black hole.
For small covering parameter $m$ and thus small winding numbers $w<m$, the difference between entwinement and the length of a winding geodesic is small, though increasing with $w$.
Thus, for regions within the entanglement shadow but far away from the black hole horizon the entanglement builds geometry proposal seems to be robust: to good approximation, entwinement is dual to the length of a winding geodesic.
In fact, if we increase $m$ and thus enlarge the region of spacetime probed by the winding geodesics the $1/N$ corrections become even smaller due to the exponential suppression with $m$.
The increase with $w$ for fixed $m$ also fits in the expectation that the length of geodesics with larger winding number should be less well approximated by entwinement.
We expect similar conclusions to hold for the two-sided black hole where the winding geodesics can probe the black hole interior.

For large winding numbers, our results of \secref{sec:large-winding-numbers} are less universal.
They depend on the structure of the seed theory of the $S_N$ orbifold.
This is not unexpected.
If we probe very deep into the interior or look at very fine-grained entanglement structure in the CFT, we might expect to see effects that depend on details of the bulk UV completion.
Moreover, the results of \secref{sec:large-winding-numbers} are valid only at the orbifold point where the notion of a classical geometry is not applicable to the description of the bulk physics.
For that reason, we cannot draw any conclusion about the validity of ``entanglement builds geometry'' for regions close to the black hole horizon.
By regions close to the black hole horizon we mean more specifically points in the entanglement shadow whose proper distance to the horizon is exponentially small in $l/G_N$, where $l$ is the AdS radius.
These regions are probed only by geodesics with large winding numbers $w$ scaling proportional to $l/G_N$.\footnote{The turning point for a geodesic with opening angle $2\pi\Delta\phi = 2\pi(L+w)$ is located at $r=r_0\coth\left(\frac{r_0\Delta\phi}{2l}\right)$ in BTZ coordinates $ds^2=-\frac{r^2-r_0^2}{l^2}dt^2 + \frac{l^2}{r^2-r_0^2}dr^2 + r^2d\phi^2$. The turning point is the point where the geodesic reaches deepest into the bulk. The proper distance between the turning point and the horizon $r=r_0$ is given by $l \log\left(\coth\left(\frac{r_0\Delta\phi}{4l}\right)\right)$ which can be approximated as $2l \exp\left(-\frac{\pi l}{\beta}\Delta\phi\right)$ for large $\Delta\phi$.}
However, we can draw a negative conclusion on whether entwinement being well approximated by a geodesic length can serve as an indicator for a semiclassical bulk description being applicable.
Given that the corrections computed in \secref{sec:large-winding-numbers} are small, it is clear that entwinement can ``look geometrically'' in the sense of being well approximated by geodesic length even though the dual bulk physics is beyond the applicability of semiclassical physics.

In summary, our computations imply that the ``entanglement builds geometry'' proposal is robust for regions inside the entanglement shadow of a one-sided BTZ black hole sufficiently far away from the horizon. For regions close to the horizon our calculations can not probe the validity of ``entanglement builds geometry''. However, our results imply that even though the bulk physics is not well approximated by a semiclassical description for this region, the entanglement entropy or entwinement is nevertheless well approximated by a geodesic length.

\medskip

Using conformal perturbation theory, it is possible to move away from the orbifold point.
The leading correction to the results of \secref{sec:large-winding-numbers} in conformal perturbation theory would be determined from the anomalous dimension of the first excited state $\ket{\psi_1}^{\otimes N}$ in the untwisted sector (at high temperatures and small gap $\Delta_1$) or of the twist sector ground state (either at low temperatures or at high temperatures and large gap $\Delta_1$).
However, a reliable conclusion about the behaviour of entwinement for large winding numbers at strong coupling in the CFT would require a much more complete understanding of the anomalous dimensions in all twisted sectors\footnote{For low temperatures, understanding the maximally twisted sector would be sufficient. But for high temperatures, a modular $S$ transformation $\tau \to -1/\tau$ exchanges space and time and thus relates the maximally twisted sector to all other twisted sectors.} which is beyond what can be done with conformal perturbation theory at the moment.

Interestingly, under reasonable assumptions detailed in \secref{sec:covering-space}, the conclusions drawn above for small winding numbers do not depend strongly on the coupling constant.
Even at weak coupling in the CFT where the strings in the dual gravity description are parametrically large and the approximation by general relativity is not expected to be good, the duality between entwinement and the length of a geodesics holds to good approximation.
That the bulk geometry is encoded in terms of entanglement in the boundary theory when GR is a good approximation to the bulk physics does not mean that entanglement can't have a geometric appearance when GR is not a good approximation.
In fact, at weak coupling we find that even for very large winding numbers entwinement can be dual to the length of a geodesic to good approximation.

Let us close with an outlook on future directions.
So far, entwinement has been exclusively studied in AdS$_3$/CFT$_2$.
It would he highly interesting to study entanglement between internal degrees of freedom also in higher dimensions.
Holographic CFTs contain generically a large number of internal degrees of freedom, whether in two dimensions or higher.
Probing entanglement between these internal degrees of freedom, but also between spatial ones, might improve our understanding of the encoding of the bulk geometry in terms of boundary degrees of freedom just as entwinement has done in AdS$_3$/CFT$_2$.
A similar research direction, which has received relatively little attention, is probing the geometry of the compact directions in string theory AdS/CFT constructions.
Here, entanglement between internal degrees of freedom might play an important role too, see \cite{Mollabashi:2014qfa,Karch:2014pma,Taylor:2015kda,Das:2022njy} for previous work in this direction.

\section*{Acknowledgments}
We would like to thank Jiuci Xu, Andrew Rolph and Christoph Uhlemann for useful discussions. Work at VUB was supported by FWO-Vlaanderen project G012222N and by the VUB Research Council through the Strategic Research Program High-Energy Physics. MG is supported by FWO-Vlaanderen through a Junior Postdoctoral Fellowship 1238224N. DH is supported by a PhD fellowship from the VUB Research Council.

\appendix
\section{Details of the covering space interpretation}
\label{app:covering-space}
In \secref{sec:covering-space}, we discussed the difference between the thermal partition function on the covering space and the partition function for the subset of twisted sectors which is of interest for entwinement.
In particular, we explained how to rewrite the dominant Virasoro character for the latter partition function as a combination of covering space Virasoro characters in order to determine which bulk fields need to be taken into account to compute $1/N$ corrections for entwinement.
In this appendix, we briefly explain that the same rewriting can be done for the entire partition function at the orbifold point instead of just the dominant character.
This provides a non-perturbative (in $G_N$) correspondence between entwinement and ordinary entanglement on the covering space.

We will start with a simple example where $N=m=2$ and work out the computation relevant for the high temperature limit of entwinement, where the argument of the characters is related by a modular transformation to the standard form.
In the case $N=m=2$, the subset of twisted sectors is given by the $(2)$ sector with partition function
\begin{equation}
  Z^{(2)}_{\text{subset},2}(\tau) = Z_{(2)}(\tau) = \frac{1}{2}\left[\tilde Z\left(\frac{\tau}{2}\right) + \tilde Z\left(\frac{\tau+1}{2}\right)\right],
\end{equation}
while the covering space partition function is given by
\begin{equation}
  Z^{(2)}_{\text{subset},2}(\tau) = \tilde Z\left(\frac{\tau}{2}\right).
\end{equation}
By using modular invariance of the seed partition function $\tilde Z$, we can rewrite this in a form more suited for the high temperature expansion we are interested in,
\begin{equation}
  Z_{(2)}(\tau) = \frac{1}{2}\left[\tilde Z\left(-\frac{2}{\tau}\right) + \tilde Z\left(-\frac{1}{2\tau}+\frac{1}{2}\right)\right].
  \label{eq:example-partition-function-subset}
\end{equation}
Inserting the expansion into characters, we get
\begin{equation}
  Z_{(2)}(\tau) = \frac{1}{2}\sum_{(h,\bar h)}\left[\chi_h^{(c/2)}\left(-\frac{2}{\tau}\right)\chi_{\bar h}^{(c/2)}\left(-\frac{2}{\tau}\right) + \chi_h^{(c/2)}\left(-\frac{1}{2\tau}+\frac{1}{2}\right)\chi_{\bar h}^{(c/2)}\left(-\frac{1}{2\tau}+\frac{1}{2}\right)\right].
  \label{eq:example-partition-function-subset2}  
\end{equation}
We would now like to rewrite this as a single expansion into characters $\chi_h^{(c/2)}(-2/\tau)$,
\begin{equation}
  Z_{(2)}(\tau) = \sum_{(h,\bar h)} a_{h,\bar{h}} \chi_h^{(c/2)}\left(-\frac{2}{\tau}\right)\chi_{\bar h}^{(c/2)}\left(-\frac{2}{\tau}\right),
  \label{eq:rewriting-example-partition-function-subset}
\end{equation}
as was done in the main text for the dominant character at large $N$.
The first term in \eqref{eq:example-partition-function-subset2} is already of the form we want.
To deal with the second term in \eqref{eq:example-partition-function-subset2}, we expand in Virasoro characters and rewrite the $1/\left|\eta\left(-\frac{1}{2\tau}+\frac{1}{2}\right)\right|^2$ factors in those characters into a series in ${\tilde q},\bar {\tilde q}$ times a $1/\left|\eta(-2/\tau)\right|^2$ factor.
The first step of this computation is to do this rewriting for the sum $1/{\left|\eta\left(-\frac{1}{2\tau}\right)\right|^2} + 1/{\left|\eta\left(-\frac{1}{2\tau}+\frac{1}{2}\right)\right|^2}$.
First, note that
\begin{equation}
    ({\tilde q} \bar {\tilde q})^{-\frac{1}{48}}\frac{1}{2}\left(\frac{1}{\left|\eta\left(-\frac{1}{2\tau}\right)\right|^2} + \frac{1}{\left|\eta\left(-\frac{1}{2\tau}+\frac{1}{2}\right)\right|^2}\right) = \sum_{k,\bar k=0}^\infty p(k)p(\bar k){\tilde q}^{k/2}\bar {\tilde q}^{\bar k/2}\delta_{k-\bar k \in 2\ZZ}.
\end{equation}
We then perform the following resummation,
\begin{equation}
    \sum_{k,\bar k=0}^\infty p(k)p(\bar k){\tilde q}^{k/2}\bar {\tilde q}^{\bar k/2}\delta_{k-\bar k \in 2\ZZ} = \sum_{k,\bar k=0}^\infty \begin{aligned}[t]
        \biggl[&\sum_{l=0}^1 p(4k+2+l)p(4\bar k+l){\tilde q}^{2k+1+l/2}\bar {\tilde q}^{2\bar k+l/2}\\
        +&\sum_{l=0}^1 p(4k+l)p(4\bar k+2+l){\tilde q}^{2k+l/2} \bar {\tilde q}^{2\bar k+1+l/2}\\
        +&\sum_{l=0}^3 p(4k+l)p(4\bar k+l){\tilde q}^{2k+l/2}\bar {\tilde q}^{2\bar k+l/2}
    \biggr]. \end{aligned}
\end{equation}
Using \eqref{eq:relation-for-rewriting-eta} gives
\begin{equation}
    \sum_{k,\bar k=0}^\infty p(k)p(\bar k){\tilde q}^{k/2}\bar {\tilde q}^{\bar k/2}\delta_{k-\bar k \in 2\ZZ} = \frac{({\tilde q}\bar {\tilde q})^{\frac{1}{12}}}{\left|\eta(-2/\tau)\right|^2}\sum_{k,\bar k=0}^\infty \begin{aligned}[t]
        \biggl[&\sum_{l=0}^1 p_4(4k+2+l)p_4(4\bar k+l){\tilde q}^{2k+1+l/2}\bar {\tilde q}^{2\bar k+l/2}\\
        +&\sum_{l=0}^1 p_4(4k+l)p_4(4\bar k+2+l){\tilde q}^{2k+l/2} \bar {\tilde q}^{2\bar k+1+l/2}\\
        +&\sum_{l=0}^3 p_4(4k+l)p_4(4\bar k+l){\tilde q}^{2k+l/2}\bar {\tilde q}^{2\bar k+l/2}
    \biggr]. \end{aligned}
    \label{eq:rewriting-eta-eq-1}
\end{equation}
We can use the same methods to rewrite
\begin{equation}
    \frac{({\tilde q}\bar {\tilde q})^{\frac{1}{48}}}{\left|\eta\left(-\frac{1}{2\tau}\right)\right|^2} = \frac{({\tilde q}\bar {\tilde q})^{\frac{1}{12}}}{\left|\eta\left(-\frac{2}{\tau}\right)\right|^2} \sum_{k,\bar k=0}^\infty \sum_{l,\bar l = 0}^3 p_4(4k+l)p_4(4\bar k+\bar l){\tilde q}^{2k+l/2}\bar {\tilde q}^{2\bar k+\bar l/2}.
    \label{eq:rewriting-eta-eq-2}
\end{equation}
Putting \eqref{eq:rewriting-eta-eq-1} and \eqref{eq:rewriting-eta-eq-2} together, we find
\begin{equation}
    \frac{1}{\left|\eta\left(-\frac{1}{2\tau}+\frac{1}{2}\right)\right|^2} = \frac{({\tilde q}\bar {\tilde q})^{\frac{1}{24}\left(2-\frac{1}{2}\right)}}{\left|\eta\left(-\frac{2}{\tau}\right)\right|^2}\sum_{k,\bar k=0}^\infty \sum_{l,\bar l=0}^3 p_4(4k+l)p_4(4\bar k+\bar l)(2\delta_{l \bar l}+2\delta_{l,\bar l+2}+2\delta_{l+2,\bar l}-1){\tilde q}^{2k+l/2}\bar {\tilde q}^{2\bar k+\bar l/2}.
    \label{eq:rewriting-eta-eq-3}
\end{equation}
Comparing with \eqref{eq:rewriting-example-partition-function-subset}, the $p_4(4k+l)p_4(4\bar k+\bar l)(2\delta_{l \bar l}+2\delta_{l,\bar l+2}+2\delta_{l+2,\bar l}-1)$ factor turns into the multiplicity factor $a_{h,\bar h}$.

Going beyond the example considered above, one can see that similar expansions of the partition function $Z^{(m)}_{\text{subset},N}$ into (modular transformed) covering space characters work for any $N$ and $m$.
$Z^{(m)}_{\text{subset},N}$ contains terms with fractional powers of $\tilde q$ from various combinations of Dedekind eta functions in the Virasoro characters of seed partitions functions from the expansion in \eqref{eq:projected-partition-function-orbifold}.
We can write these fractional powers as ${\tilde q}^{k m/j^2}$ where $j$ divides $m$ and $k$ is a summation variable from the series expansion of the eta functions in $\tilde q$.
The summation over $k$ can always be split up into summations of $0 \leq l < j^2$ and $\hat k$ by defining $k = j^2 \hat k + l$.
Restrictions on the allowed spin of descendants can be incorporated into this splitting by restricting the summation range for $l$ and its antiholomorphic counterpart $\bar l$, for instance by inserting Kronecker deltas similar to \eqref{eq:rewriting-eta-eq-3}.
What remains is a series in ${\tilde q}^{k m/h^2} = {\tilde q}^{m\hat k}q^{m l/j^2}$ with appropriate coefficients.
Using \eqref{eq:relation-for-rewriting-eta}, this series can be combined into a $1/|\eta(-m/\tau)|^2$ prefactor times a remaining series in $\tilde q$.
The exponents of $\tilde q$ that turn up in this remaining series determine the ``primary spectrum'' and the coefficients thereof the ``multiplicity factors'' that are used as input in the bulk computation of the $1/N$ corrections.
Thus, barring practical difficulties in determining the spectrum and multiplicity, it is possible to exactly rewrite the subset partition function in order to relate entwinement to an ordinary entanglement entropy in a covering theory.

\section{$1/N$ corrections for large intervals}
\label{app:large-intervals}
Just as for the ordinary entanglement entropy, entwinement undergoes a phase transition as the size of the interval $[0,(L+w)/m]$ in the covering space is varied \cite{Gerbershagen:2021gvc}.
In the main text, only small intervals were considered.
This appendix contains computations of the $1/N$ corrections for large intervals using the monodromy method which we already employed in \secref{sec:monodromy-method}.

For large $w$, the monodromy condition that gives the dominant conformal block is given by trivial monodromy around a) $n$ times the time circle and b) the complement of the entangling interval plus $m-w-1$ times the spatial circle.
In this case, the high temperature limit of interest is ${\tilde q} \to 0$ while keeping $\tilde u_2 = e^{-2\pi i (m-z_2)/\tau} = e^{-2\pi i (m-w-L)/\tau}$ fixed.
The uniformization equation is given by
\begin{equation}
  \tilde\Psi''(u) + \biggl[
  \begin{aligned}[t]
    &\frac{\Delta}{2m}\sum_k \frac{{\tilde q}^k}{u(u-{\tilde q}^k)^2} + \frac{\Delta}{2m} \sum_k \frac{{\tilde q}^k/\tilde u_2}{u(u-{\tilde q}^k/\tilde u_2)^2}\\
    &+ \left(\frac{\Delta}{2m} + \tilde u_2 \partial_{\tilde u_2} \tilde f_\text{cl.}\right) \sum_k \frac{{\tilde q}^k({\tilde q}^m/\tilde u_2-1)}{u(u-{\tilde q}^k)(u-{\tilde q}^{k+m}/\tilde u_2)} + \frac{1/4 - {\tilde q} \partial_{\tilde q} \tilde f_\text{cl.} - m\tilde u_2 \partial_{\tilde u_2}\tilde f_\text{cl.}}{u^2}
  \biggr]\tilde\Psi(u) = 0.
  \end{aligned}
\end{equation}
We demand trivial monodromy for $u \to u e^{2\pi i n}$.
To simplify imposing this monodromy condition, let us perform another coordinate transformation
\begin{equation}
    v = u^{1/n}, \quad \tilde \Psi(u) = \left(\frac{\partial v}{\partial u}\right)^{h_\Psi}\hat\Psi(v), \quad \tilde f_\text{cl.} = \hat f_\text{cl.} - \sum_{i=1,2} h_i \log\left(\frac{\partial v(u_i)}{\partial u_i}\right).
\end{equation}
This gives
\begin{equation}
    \hat\Psi''(v) + \biggl[
    \begin{aligned}[t]
        &\frac{\Delta n^2}{2m}\sum_k \frac{{\tilde q}^k v^n}{v^2(v^n-{\tilde q}^k)^2} + \frac{\Delta n^2}{2m} \sum_k \frac{{\tilde q}^k v^n/\tilde v_2^n}{v^2(v^n-{\tilde q}^k/\tilde v_2^n)^2}\\
        &+ n\left(\frac{\Delta}{2m} + \tilde v_2 \partial_{\tilde v_2} \hat f_\text{cl.}\right) \sum_k \frac{{\tilde q}^k v^n({\tilde q}^m/\tilde v_2^n-1)}{v^2(v^n-{\tilde q}^k)(v^n-{\tilde q}^{k+m}/\tilde v_2^n)}\\
        &+ n^2\frac{1/4 - {\tilde q} \partial_{\tilde q} \hat f_\text{cl.} - nm\tilde v_2 \partial_{\tilde v_2}\hat f_\text{cl.} - \frac{\Delta}{2}n(1-n)}{v^2}
  \biggr]\hat\Psi(v) = 0.
  \end{aligned}
\end{equation}
Solving this equation in a perturbation expansion is more difficult, as the equation does not reduce to the case of a single interval on the $v$ plane as ${\tilde q} \to 0$,
Instead, we get $n$ intervals distributed in a $\ZZ_n$ symmetric fashion:
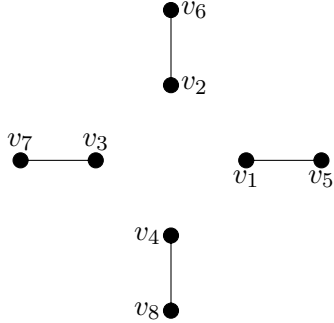
\begin{center}
    \begin{tikzpicture}
        \filldraw (1,0) circle[radius=0.1] node[below] {$v_1$} -- (2,0) circle[radius=0.1] node[below] {$v_5$};
        \filldraw (0,1) circle[radius=0.1] node[right] {$v_2$} -- (0,2) circle[radius=0.1] node[right] {$v_6$};
        \filldraw (-1,0) circle[radius=0.1] node[above] {$v_3$} -- (-2,0) circle[radius=0.1] node[above] {$v_7$};
        \filldraw (0,-1) circle[radius=0.1] node[left] {$v_4$} -- (0,-2) circle[radius=0.1]     node[left] {$v_8$};
    \end{tikzpicture}    
\end{center}
To see, this note that the uniformization equation at ${\tilde q} \to 0$ reduces to
\begin{equation}
    \hat\Psi_0''(v) + \biggl[
    \begin{aligned}[t]
        &\frac{\Delta n^2}{2m}\left(\frac{v^n}{v^2(v^n-1)^2}+\frac{v^n/\tilde v_2^n}{v^2(v^n-1/\tilde v_2^n)^2}\right)\\
        &+ n\left(\frac{\Delta}{2m} + \tilde v_2 \partial_{\tilde v_2} \hat f_\text{cl.}\right)\left(\frac{1/\tilde v_2^n}{v^2(v^n-1/\tilde v_2^n)}-\frac{1}{v^2(v^n-1)}\right)\\
        &+ \frac{1/4 - n^2 {\tilde q} \partial_{\tilde q} \hat f_\text{cl.}-\frac{\Delta}{2}n(1-n)}{v^2}
  \biggr]\hat\Psi_0(v) = 0.
  \end{aligned}
  \label{eq:uniformization-equation-large-interval-high-temperature}
\end{equation}
This is precisely the uniformization equation on the plane for $n$ intervals with endpoints $v_k = e^{2\pi i (k-1)/n},v_{k+n} = e^{2\pi i (k-1)/n}/\tilde v_2$,
\begin{equation}
    \hat\Psi_0''(v) + \sum_{i=1}^{2n}\biggl[\frac{\Delta/2}{(v-v_i)^2} - \frac{\partial_{v_i}\hat f_\text{cl.}}{v-v_i}\biggr]\hat\Psi_0(v) = 0.
  \label{eq:uniformization-equation-comparison-large-interval-high-temperature}
\end{equation}
This follows from $\partial_{v_k} = e^{-2\pi i (k-1)/n}\partial_{v_{1}},\partial_{v_{k+n}} = e^{-2\pi i (k-1)/n}\partial_{v_{n+1}}$ and $\sum_i \partial_{v_i}\hat f_\text{cl.} = \sum_i (v_i \partial_{v_i}\hat f_\text{cl.} - h_i) = \sum_i (v_i^2\partial_{v_i}\hat f_\text{cl.} - 2h_i v_i)=0$.
For ${\tilde q}\partial_{\tilde q} \hat f_\text{cl.} = \frac{1}{4}$, we find that \eqref{eq:uniformization-equation-large-interval-high-temperature} is equal to \eqref{eq:uniformization-equation-comparison-large-interval-high-temperature}.
To cross-check that ${\tilde q}\partial_{\tilde q} \hat f_\text{cl.} = \frac{1}{4}$ is correct, we can use the results for $f_\text{cl.}$ expanded in $n-1$ from \cite{Gerbershagen:2021yma},
\begin{equation}
    f_\text{cl.} = - \frac{\pi i}{2n^2\tau} + (n-1)\log(\tau\sinh(\pi i(1-z_2)/\tau)) + O((n-1)^2).
\end{equation}
This is equivalent to $\hat f_\text{cl.} = \frac{1}{4n^2}\log({\tilde q}) + (n-1)\log(\tilde v_2-1) + \frac{1}{2}(n-1)\log({\tilde q}) + O((n-1)^2)$, in agreement with ${\tilde q}\partial_{\tilde q} \hat f_\text{cl.} = \frac{1}{4}$ at leading order in $n-1$.

The difficulty is now that even the ${\tilde q} \to 0$ limit of the uniformization equation, eq.~\eqref{eq:uniformization-equation-large-interval-high-temperature}, has no elementary solution.
Therefore, we need to expand further, for instance in $\tilde v_2-1$ and solve the resulting equation order by order in both ${\tilde q}$ and $\tilde v_2-1$.
We can then impose trivial monodromy around $v=0$, i.e.~for $v \to v e^{2\pi i}$, and around $v=1$.
The resulting solution $\hat\Psi_\pm(v)$ is given by a Laurent expansion in $v$ with convergence radius $1 \leq v \leq {\tilde q}^{-1/n}/\tilde v_2$, determined similarly to the argument given in sec.~8.1.~of \cite{Barrella:2013wja}.
Therefore, the region where both $\hat\Psi_\pm(v)$ and $\hat\Psi_\pm(v/{\tilde q}^{1/n})$ are convergent is given by ${\tilde q}^{-1/n} \leq v \leq {\tilde q}^{-1/n}/\tilde v_2$.
By expanding in $v$, we can thus find the monodromy around the spatial circle $v \to v/{\tilde q}^{1/n}$,
\begin{equation}
    \left(\begin{array}{c}
         \hat\Psi_+({\tilde q} v) \\
         \hat\Psi_-({\tilde q} v)
    \end{array}\right) = T \left(\begin{array}{c}
         \hat\Psi_+(v) \\
         \hat\Psi_-(v)
    \end{array}\right).
\end{equation}
From this monodromy matrix, we find one of the $n$ generators of the Schottky group.
Explicitly, we find to the first few orders in $\tilde v_2-1$ and ${\tilde q}$
\begin{equation}
    \begin{aligned}
        \hat\Psi_+(v) &= 1 - \frac{n\Delta}{12m}\left(\frac{1+(n-1)v^n}{(v^n-1)^2} + {\tilde q}\left[\frac{n-1}{v^n}+(n+1)v^n\right]\right)(\tilde v_2-1)^2 + ...\\
        \hat\Psi_-(v) &= v - \frac{n\Delta}{12m}v\left(\frac{-1+(n+1)v^n}{(v^n-1)^2} + {\tilde q}\left[\frac{n+1}{v^n}+(n-1)v^n\right]\right)(\tilde v_2-1)^2 + ...
    \end{aligned}
    \label{eq:solution-decoupling-equation-v-large-interval-monodromy}
\end{equation}
Equivalently, a solution in terms of $u$ is given by
\begin{equation}
    \begin{aligned}
        \tilde\Psi_+(u) &= u^{\frac{1}{2}\left(1-\frac{1}{n}\right)}\left[1+\frac{\Delta}{12nm}\left(\frac{u(u-(n+1))}{(u-1)^2}-{\tilde q}\left(\frac{n-1}{u}+(n+1)u\right)\right)(\tilde u_2-1)^2\right] + ...\\
        \tilde\Psi_-(u) &= u^{\frac{1}{2}\left(1+\frac{1}{n}\right)}\left[1-\frac{\Delta}{12nm}\left(\frac{u(u+(n-1))}{(u-1)^2}+{\tilde q}\left(\frac{n+1}{u}+(n-1)u\right)\right)(\tilde u_2-1)^2\right] + ...\\
    \end{aligned}
    \label{eq:solution-decoupling-equation-u-large-interval-monodromy}
\end{equation}
Because of the $u^{\frac{1}{2}\left(1 \pm \frac{1}{n}\right)}$ prefactor with fractional scaling in $u$, the monodromy matrix $T$ must be diagonal.
An instructive example is the limit $\tilde u_2 \to 1$, in which case $\tilde \Psi_\pm(u) = u^{\frac{1}{2}\left(1 \mp \frac{1}{n}\right)}$.
In this limit, the monodromy matrix is given by
\begin{equation}
    \left(\begin{array}{c}
         \tilde\Psi_+({\tilde q} u) \\
         \tilde\Psi_-({\tilde q} u)
    \end{array}\right) = T \left(\begin{array}{c}
         \tilde\Psi_+(u) \\
         \tilde\Psi_-(u)
    \end{array}\right) \quad \Leftrightarrow \quad T/\sqrt{\det(T)} = \left(\begin{array}{cc}
       {\tilde q}^{\frac{1}{2n}}  & 0 \\
       0 & {\tilde q}^{-\frac{1}{2n}}.
    \end{array}\right)
\end{equation}
We have normalized the matrix by $1/\sqrt{\det(T)}$ in order to obtain an $SL(2,\CC)$ element (instead of $GL(2,\CC)$), which can always be achieved by rescaling the solutions of the uniformization equation by a constant prefactor.
Then we can immediately read off the smallest eigenvalue $Q_\Gamma = {\tilde q}^{\frac{1}{n}}$.
Let us check that we obtain the thermal entropy in the $\tilde u_2 \to 1$ limit.
Using the formula \eqref{eq:1-over-N-correction-partition-function} for the partition function and the definition of the Rényi entropy
\begin{equation}
    S_n = -\frac{1}{n-1}(\log Z_n - n \log Z_1)
\end{equation}
we find the change in entropy
\begin{equation}
    \delta S(\beta) = \sum_{m=2}^\infty\left(\frac{m {\tilde q}^m \log {\tilde q}}{1-{\tilde q}^m} - \log(1-{\tilde q}^m)\right).
    \label{eq:1-over-N-correction-thermal-entropy}
\end{equation}
This can be compared with the thermal entropy arising from the character of the Virasoro group
\begin{equation}
    S(\beta) = - \beta^2\partial_\beta(\beta^{-1}\log\chi_0^{(c)}(2\pi/\beta)) = \frac{c}{3}\frac{\pi^2}{\beta} + \delta S(\beta)
\end{equation}
where the vacuum character is given by $\chi_0^{(c)}(2\pi/\beta) = \frac{{\tilde q}^{-\frac{c-1}{24}}(1-{\tilde q})}{\eta(2\pi i/\beta)} = {\tilde q}^{-\frac{c}{24}}/\prod_{m=2}^\infty(1-{\tilde q}^m)$.
As expected, the subleading order of the thermal entropy at large central charge matches with the entropy obtained from the monodromy method in the limit that the boundary interval covers the entire space and the RT surface localizes to the black hole horizon.
For entwinement monodromy conditions, we need to compute the monodromy around an $m$ times larger spatial circle.
The monodromy matrix for that is just given by $T^m$.
The change in entropy in that case is just given by \eqref{eq:1-over-N-correction-thermal-entropy} with $\beta \to \beta/m$, corresponding to the thermal entropy arising from $\chi_0^{(c/m)}(\beta/m)$.

In order to find the remaining $n-1$ Schottky generators, we need to determine the monodromy around paths which encircle the endpoints $u=1,1/\tilde u_2$ of the complement of the entangling interval a number of times to get to a different sheet of the Riemann surface.
This monodromy cannot be read off from the solution \eqref{eq:solution-decoupling-equation-u-large-interval-monodromy} because at zeroth order in the $\tilde u_2-1$ perturbation expansion, the two endpoints $\tilde u_2$ and $1$ are the same so that we can only determine the monodromy around both points in the perturbation expansion used in \eqref{eq:solution-decoupling-equation-u-large-interval-monodromy}.
Therefore, we perform a coordinate transformation $w=\frac{1}{v}\frac{v-1}{1-\tilde v_2}$ that maps the interval endpoints to $w=0,1$ and $v=0$ to $w=\infty$.
In these coordinates, we expand the uniformization equation again in ${\tilde q}$ and $\tilde v_2-1$, solving order by order and demanding trivial monodromy around $w=0,1$ as well as for $w \to w e^{2\pi i}$.
This gives
\begin{equation}
    \begin{aligned}
    \bar\Psi_+(w) &= w^{\frac{1}{2}\left(1+\sqrt{1-2\Delta/m}\right)}(w-1)^{\frac{1}{2}\left(1-\sqrt{1-2\Delta/m}\right)} + ...\\
    \bar\Psi_-(w) &= w^{\frac{1}{2}\left(1-\sqrt{1-2\Delta/m}\right)}(w-1)^{\frac{1}{2}\left(1+\sqrt{1-2\Delta/m}\right)} + ...
    \end{aligned}
    \label{eq:solution-decoupling-equation-w-large-interval-monodromy}
\end{equation}
so that the monodromy around $w=0$ and around $w=1$ diagonalizes,
\begin{equation}
    M_0 = \left(\begin{array}{cc}
    e^{\pi i\left(1+\sqrt{1-2\Delta/m}\right)} & 0\\
    0 & e^{\pi i\left(1-\sqrt{1-2\Delta/m}\right)}\end{array}\right),
    \qquad
    M_1 = \left(\begin{array}{cc}
    e^{\pi i\left(1-\sqrt{1-2\Delta/m}\right)} & 0\\
    0 & e^{\pi i\left(1+\sqrt{1-2\Delta/m}\right)}\end{array}\right).
\end{equation}
Finally, we need to match the two solutions \eqref{eq:solution-decoupling-equation-w-large-interval-monodromy} and \eqref{eq:solution-decoupling-equation-v-large-interval-monodromy} in the subregion of the complex plane where both are valid.
As $\hat\Psi_\pm(v)$ is valid for $v \gg {\tilde q},\tilde v_2-1$ and $\bar\Psi_\pm(w)$ is valid for $w \gg {\tilde q},\tilde v_2-1$, we can match the two solutions for $v \gg 1$.
Concretely, we make an ansatz
\begin{equation}
    \left(\begin{array}{c}
         \bar\Psi_+(w)\\
         \bar\Psi_-(w)\end{array}\right)
    =
    U \left(\frac{\partial v}{\partial w}\right)^{h_\psi}
    \left(\begin{array}{c}
         \hat\Psi_+(v)\\
         \hat\Psi_-(v) 
    \end{array}\right)
\end{equation}
and expand in ${\tilde q},\tilde v_2-1$ and $1/v$, determining the matrix $U$ in a series in ${\tilde q}$ and $\tilde v_2-1$.
Finally, the Schottky generators are given by
\begin{equation}
    L_k = M_1^{k-1} U T^m U^{-1} M_1^{-(k-1)}, \quad 1 \leq k \leq n.
\end{equation}
Following the same steps as for small intervals, we find the leading $1/N$ correction to entwinement from bulk metric fluctuations to be given by
\begin{equation}
    \begin{aligned}
        \delta S_{m,w}(L) = 2{\tilde q}^{2m} \left(1+2m \log({\tilde q})-\frac{2}{3}(\tilde u_2-1)^2+\frac{2}{3}(\tilde u_2-1)^3 + O((\tilde u_2-1)^4)\right) + O({\tilde q}^{4m}).
    \end{aligned}
    \label{eq:1-over-N-corrections-large-intervals}
\end{equation}
where $\tilde u_2 = e^{-\frac{4\pi^2}{\beta}(m-w-L)}$ and ${\tilde q}=e^{-\frac{4\pi^2}{\beta}}$.
Higher order terms in the $\tilde q$ expansion are determined from Schottky group elements built out of multi-letter words.
For the first order in the $\tilde q$ expansion, the $1/N$ corrections for large intervals are given by the sum of the $1/N$ corrections for the thermal entropy and the $1/N$ corrections for small intervals with the replacement of the interval in the covering space by its complement, $\delta S_{m,w}(L) = \delta S(\beta) + \delta S_{m,m-w-1}(1-L) + O(\tilde q^{4m})$.

\bibliographystyle{JHEP}
\bibliography{bibliography.bib}

\providecommand{\href}[2]{#2}\begingroup\raggedright\begin{thebibliography}{10}

\bibitem{Swingle:2009bg}
B.~Swingle, \emph{{Entanglement Renormalization and Holography}},
  \href{https://doi.org/10.1103/PhysRevD.86.065007}{\emph{Phys. Rev. D}
  {\bfseries 86} (2012) 065007}
  [\href{https://arxiv.org/abs/0905.1317}{{\ttfamily 0905.1317}}].

\bibitem{VanRaamsdonk:2010pw}
M.~Van~Raamsdonk, \emph{{Building up spacetime with quantum entanglement}},
  \href{https://doi.org/10.1007/s10714-010-1034-0}{\emph{Gen. Rel. Grav.}
  {\bfseries 42} (2010) 2323}
  [\href{https://arxiv.org/abs/1005.3035}{{\ttfamily 1005.3035}}].

\bibitem{Ryu:2006bv}
S.~Ryu and T.~Takayanagi, \emph{{Holographic derivation of entanglement entropy
  from AdS/CFT}},
  \href{https://doi.org/10.1103/PhysRevLett.96.181602}{\emph{Phys. Rev. Lett.}
  {\bfseries 96} (2006) 181602}
  [\href{https://arxiv.org/abs/hep-th/0603001}{{\ttfamily hep-th/0603001}}].

\bibitem{Faulkner:2013ana}
T.~Faulkner, A.~Lewkowycz and J.~Maldacena, \emph{{Quantum corrections to
  holographic entanglement entropy}},
  \href{https://doi.org/10.1007/JHEP11(2013)074}{\emph{JHEP} {\bfseries 11}
  (2013) 074} [\href{https://arxiv.org/abs/1307.2892}{{\ttfamily 1307.2892}}].

\bibitem{Engelhardt:2014gca}
N.~Engelhardt and A.C.~Wall, \emph{{Quantum Extremal Surfaces: Holographic
  Entanglement Entropy beyond the Classical Regime}},
  \href{https://doi.org/10.1007/JHEP01(2015)073}{\emph{JHEP} {\bfseries 01}
  (2015) 073} [\href{https://arxiv.org/abs/1408.3203}{{\ttfamily 1408.3203}}].

\bibitem{Balasubramanian:2014sra}
V.~Balasubramanian, B.D.~Chowdhury, B.~Czech and J.~{de Boer},
  \emph{{Entwinement and the emergence of spacetime}},
  \href{https://doi.org/10.1007/JHEP01(2015)048}{\emph{JHEP} {\bfseries 01}
  (2015) 048} [\href{https://arxiv.org/abs/1406.5859}{{\ttfamily 1406.5859}}].

\bibitem{Balasubramanian:2016xho}
V.~Balasubramanian, A.~Bernamonti, B.~Craps, T.~De~Jonckheere and F.~Galli,
  \emph{{Entwinement in discretely gauged theories}},
  \href{https://doi.org/10.1007/JHEP12(2016)094}{\emph{JHEP} {\bfseries 12}
  (2016) 094} [\href{https://arxiv.org/abs/1609.03991}{{\ttfamily
  1609.03991}}].

\bibitem{Balasubramanian:2018ajb}
V.~Balasubramanian, B.~Craps, T.~De~Jonckheere and G.~Sárosi,
  \emph{{Entanglement versus entwinement in symmetric product orbifolds}},
  \href{https://doi.org/10.1007/JHEP01(2019)190}{\emph{JHEP} {\bfseries 01}
  (2019) 190} [\href{https://arxiv.org/abs/1806.02871}{{\ttfamily
  1806.02871}}].

\bibitem{Erdmenger:2019lzr}
J.~Erdmenger and M.~Gerbershagen, \emph{{Entwinement as a possible alternative
  to complexity}}, \href{https://doi.org/10.1007/JHEP03(2020)082}{\emph{JHEP}
  {\bfseries 03} (2020) 082}
  [\href{https://arxiv.org/abs/1910.05352}{{\ttfamily 1910.05352}}].

\bibitem{Gerbershagen:2021gvc}
M.~Gerbershagen, \emph{{Illuminating entanglement shadows of BTZ black holes by
  a generalized entanglement measure}},
  \href{https://doi.org/10.1007/JHEP10(2021)187}{\emph{JHEP} {\bfseries 10}
  (2021) 187} [\href{https://arxiv.org/abs/2105.01097}{{\ttfamily
  2105.01097}}].

\bibitem{Craps:2022pke}
B.~Craps, M.~De~Clerck and A.~Vilar~L\'opez, \emph{{Definitions of
  entwinement}}, \href{https://doi.org/10.1007/JHEP03(2023)079}{\emph{JHEP}
  {\bfseries 03} (2023) 079}
  [\href{https://arxiv.org/abs/2211.17253}{{\ttfamily 2211.17253}}].

\bibitem{Czech:2012bh}
B.~Czech, J.L.~Karczmarek, F.~Nogueira and M.~Van~Raamsdonk, \emph{{The Gravity
  Dual of a Density Matrix}},
  \href{https://doi.org/10.1088/0264-9381/29/15/155009}{\emph{Class. Quant.
  Grav.} {\bfseries 29} (2012) 155009}
  [\href{https://arxiv.org/abs/1204.1330}{{\ttfamily 1204.1330}}].

\bibitem{Hubeny:2013gta}
V.E.~Hubeny, H.~Maxfield, M.~Rangamani and E.~Tonni, \emph{{Holographic
  entanglement plateaux}},
  \href{https://doi.org/10.1007/JHEP08(2013)092}{\emph{JHEP} {\bfseries 08}
  (2013) 092} [\href{https://arxiv.org/abs/1306.4004}{{\ttfamily 1306.4004}}].

\bibitem{Nogueira:2013if}
F.~Nogueira, \emph{{Extremal Surfaces in Asymptotically AdS Charged Boson Stars
  Backgrounds}}, \href{https://doi.org/10.1103/PhysRevD.87.106006}{\emph{Phys.
  Rev. D} {\bfseries 87} (2013) 106006}
  [\href{https://arxiv.org/abs/1301.4316}{{\ttfamily 1301.4316}}].

\bibitem{Freivogel:2014lja}
B.~Freivogel, R.~Jefferson, L.~Kabir, B.~Mosk and I.-S.~Yang, \emph{{Casting
  Shadows on Holographic Reconstruction}},
  \href{https://doi.org/10.1103/PhysRevD.91.086013}{\emph{Phys. Rev. D}
  {\bfseries 91} (2015) 086013}
  [\href{https://arxiv.org/abs/1412.5175}{{\ttfamily 1412.5175}}].

\bibitem{Buividovich:2008gq}
P.V.~Buividovich and M.I.~Polikarpov, \emph{{Entanglement entropy in gauge
  theories and the holographic principle for electric strings}},
  \href{https://doi.org/10.1016/j.physletb.2008.10.032}{\emph{Phys. Lett. B}
  {\bfseries 670} (2008) 141}
  [\href{https://arxiv.org/abs/0806.3376}{{\ttfamily 0806.3376}}].

\bibitem{Donnelly:2011hn}
W.~Donnelly, \emph{{Decomposition of entanglement entropy in lattice gauge
  theory}}, \href{https://doi.org/10.1103/PhysRevD.85.085004}{\emph{Phys. Rev.
  D} {\bfseries 85} (2012) 085004}
  [\href{https://arxiv.org/abs/1109.0036}{{\ttfamily 1109.0036}}].

\bibitem{Casini:2013rba}
H.~Casini, M.~Huerta and J.A.~Rosabal, \emph{{Remarks on entanglement entropy
  for gauge fields}},
  \href{https://doi.org/10.1103/PhysRevD.89.085012}{\emph{Phys. Rev. D}
  {\bfseries 89} (2014) 085012}
  [\href{https://arxiv.org/abs/1312.1183}{{\ttfamily 1312.1183}}].

\bibitem{Radicevic:2014kqa}
D.~Radicevic, \emph{{Notes on Entanglement in Abelian Gauge Theories}},
  \href{https://arxiv.org/abs/1404.1391}{{\ttfamily 1404.1391}}.

\bibitem{Aoki:2015bsa}
S.~Aoki, T.~Iritani, M.~Nozaki, T.~Numasawa, N.~Shiba and H.~Tasaki, \emph{{On
  the definition of entanglement entropy in lattice gauge theories}},
  \href{https://doi.org/10.1007/JHEP06(2015)187}{\emph{JHEP} {\bfseries 06}
  (2015) 187} [\href{https://arxiv.org/abs/1502.04267}{{\ttfamily
  1502.04267}}].

\bibitem{Ghosh:2015iwa}
S.~Ghosh, R.M.~Soni and S.P.~Trivedi, \emph{{On The Entanglement Entropy For
  Gauge Theories}}, \href{https://doi.org/10.1007/JHEP09(2015)069}{\emph{JHEP}
  {\bfseries 09} (2015) 069}
  [\href{https://arxiv.org/abs/1501.02593}{{\ttfamily 1501.02593}}].

\bibitem{Soni:2015yga}
R.M.~Soni and S.P.~Trivedi, \emph{{Aspects of Entanglement Entropy for Gauge
  Theories}}, \href{https://doi.org/10.1007/JHEP01(2016)136}{\emph{JHEP}
  {\bfseries 01} (2016) 136}
  [\href{https://arxiv.org/abs/1510.07455}{{\ttfamily 1510.07455}}].

\bibitem{Eberhardt:2021vsx}
L.~Eberhardt, \emph{{A perturbative CFT dual for pure NS\textendash{}NS
  AdS$_{3}$ strings}}, \href{https://doi.org/10.1088/1751-8121/ac47b2}{\emph{J.
  Phys. A} {\bfseries 55} (2022) 064001}
  [\href{https://arxiv.org/abs/2110.07535}{{\ttfamily 2110.07535}}].

\bibitem{Knighton:2024qxd}
B.~Knighton and V.~Sriprachyakul, \emph{{Unravelling AdS$_{3}$/CFT$_{2}$ near
  the boundary}}, \href{https://doi.org/10.1007/JHEP01(2025)042}{\emph{JHEP}
  {\bfseries 01} (2025) 042}
  [\href{https://arxiv.org/abs/2404.07296}{{\ttfamily 2404.07296}}].

\bibitem{Sriprachyakul:2024gyl}
V.~Sriprachyakul, \emph{{Superstrings near the conformal boundary of
  AdS$_{3}$}}, \href{https://doi.org/10.1007/JHEP08(2024)203}{\emph{JHEP}
  {\bfseries 08} (2024) 203}
  [\href{https://arxiv.org/abs/2405.03678}{{\ttfamily 2405.03678}}].

\bibitem{David:2002wn}
J.R.~David, G.~Mandal and S.R.~Wadia, \emph{{Microscopic formulation of black
  holes in string theory}},
  \href{https://doi.org/10.1016/S0370-1573(02)00271-5}{\emph{Phys. Rept.}
  {\bfseries 369} (2002) 549}
  [\href{https://arxiv.org/abs/hep-th/0203048}{{\ttfamily hep-th/0203048}}].

\bibitem{Datta:2017ert}
S.~Datta, L.~Eberhardt and M.R.~Gaberdiel, \emph{{Stringy $\mathcal{N}=(2,2)$
  holography for AdS${_3}$}},
  \href{https://doi.org/10.1007/JHEP01(2018)146}{\emph{JHEP} {\bfseries 01}
  (2018) 146} [\href{https://arxiv.org/abs/1709.06393}{{\ttfamily
  1709.06393}}].

\bibitem{Martinec:2022ofs}
E.J.~Martinec, \emph{{A defect in AdS$_{3}$/CFT$_{2}$ duality}},
  \href{https://doi.org/10.1007/JHEP06(2022)024}{\emph{JHEP} {\bfseries 06}
  (2022) 024} [\href{https://arxiv.org/abs/2201.04218}{{\ttfamily
  2201.04218}}].

\bibitem{Hartman:2014oaa}
T.~Hartman, C.A.~Keller and B.~Stoica, \emph{{Universal Spectrum of 2d
  Conformal Field Theory in the Large c Limit}},
  \href{https://doi.org/10.1007/JHEP09(2014)118}{\emph{JHEP} {\bfseries 09}
  (2014) 118} [\href{https://arxiv.org/abs/1405.5137}{{\ttfamily 1405.5137}}].

\bibitem{Hartman:2013mia}
T.~Hartman, \emph{{Entanglement Entropy at Large Central Charge}},
  \href{https://arxiv.org/abs/1303.6955}{{\ttfamily 1303.6955}}.

\bibitem{Belin:2017nze}
A.~Belin, C.A.~Keller and I.G.~Zadeh, \emph{{Genus two partition functions and
  R\'enyi entropies of large c conformal field theories}},
  \href{https://doi.org/10.1088/1751-8121/aa8a11}{\emph{J. Phys. A} {\bfseries
  50} (2017) 435401} [\href{https://arxiv.org/abs/1704.08250}{{\ttfamily
  1704.08250}}].

\bibitem{Gerbershagen:2021yma}
M.~Gerbershagen, \emph{{Monodromy methods for torus conformal blocks and
  entanglement entropy at large central charge}},
  \href{https://doi.org/10.1007/JHEP08(2021)143}{\emph{JHEP} {\bfseries 08}
  (2021) 143} [\href{https://arxiv.org/abs/2101.11642}{{\ttfamily
  2101.11642}}].

\bibitem{Barrella:2013wja}
T.~Barrella, X.~Dong, S.A.~Hartnoll and V.L.~Martin, \emph{{Holographic
  entanglement beyond classical gravity}},
  \href{https://doi.org/10.1007/JHEP09(2013)109}{\emph{JHEP} {\bfseries 09}
  (2013) 109} [\href{https://arxiv.org/abs/1306.4682}{{\ttfamily 1306.4682}}].

\bibitem{Yin:2007gv}
X.~Yin, \emph{{Partition Functions of Three-Dimensional Pure Gravity}},
  \href{https://doi.org/10.4310/CNTP.2008.v2.n2.a1}{\emph{Commun. Num. Theor.
  Phys.} {\bfseries 2} (2008) 285}
  [\href{https://arxiv.org/abs/0710.2129}{{\ttfamily 0710.2129}}].

\bibitem{Giombi:2008vd}
S.~Giombi, A.~Maloney and X.~Yin, \emph{{One-loop Partition Functions of 3D
  Gravity}}, \href{https://doi.org/10.1088/1126-6708/2008/08/007}{\emph{JHEP}
  {\bfseries 08} (2008) 007} [\href{https://arxiv.org/abs/0804.1773}{{\ttfamily
  0804.1773}}].

\bibitem{NIST:DLMF}
``{NIST Digital Library of Mathematical Functions}.''

\bibitem{Calabrese:2010he}
P.~Calabrese, J.~Cardy and E.~Tonni, \emph{{Entanglement entropy of two
  disjoint intervals in conformal field theory II}},
  \href{https://doi.org/10.1088/1742-5468/2011/01/P01021}{\emph{J. Stat. Mech.}
  {\bfseries 1101} (2011) P01021}
  [\href{https://arxiv.org/abs/1011.5482}{{\ttfamily 1011.5482}}].

\bibitem{Cardy:2014jwa}
J.~Cardy and C.P.~Herzog, \emph{{Universal Thermal Corrections to Single
  Interval Entanglement Entropy for Two Dimensional Conformal Field Theories}},
  \href{https://doi.org/10.1103/PhysRevLett.112.171603}{\emph{Phys. Rev. Lett.}
  {\bfseries 112} (2014) 171603}
  [\href{https://arxiv.org/abs/1403.0578}{{\ttfamily 1403.0578}}].

\bibitem{Mollabashi:2014qfa}
A.~Mollabashi, N.~Shiba and T.~Takayanagi, \emph{{Entanglement between Two
  Interacting CFTs and Generalized Holographic Entanglement Entropy}},
  \href{https://doi.org/10.1007/JHEP04(2014)185}{\emph{JHEP} {\bfseries 04}
  (2014) 185} [\href{https://arxiv.org/abs/1403.1393}{{\ttfamily 1403.1393}}].

\bibitem{Karch:2014pma}
A.~Karch and C.F.~Uhlemann, \emph{{Holographic entanglement entropy and the
  internal space}},
  \href{https://doi.org/10.1103/PhysRevD.91.086005}{\emph{Phys. Rev. D}
  {\bfseries 91} (2015) 086005}
  [\href{https://arxiv.org/abs/1501.00003}{{\ttfamily 1501.00003}}].

\bibitem{Taylor:2015kda}
M.~Taylor, \emph{{Generalized entanglement entropy}},
  \href{https://doi.org/10.1007/JHEP07(2016)040}{\emph{JHEP} {\bfseries 07}
  (2016) 040} [\href{https://arxiv.org/abs/1507.06410}{{\ttfamily
  1507.06410}}].

\bibitem{Das:2022njy}
S.R.~Das, A.~Kaushal, G.~Mandal, K.K.~Nanda, M.H.~Radwan and S.P.~Trivedi,
  \emph{{Entanglement entropy in internal spaces and Ryu-Takayanagi surfaces}},
  \href{https://doi.org/10.1007/JHEP04(2023)141}{\emph{JHEP} {\bfseries 04}
  (2023) 141} [\href{https://arxiv.org/abs/2212.11640}{{\ttfamily
  2212.11640}}].

\end{thebibliography}\endgroup

\end{document}